# Flaring Stars in a Nontargeted Millimeter-wave Survey with SPT-3G

C. Tandoi[1], S. Guns[2], A. Foster[3], P. A. R. Ade[4], A. J. Anderson[5,6,7], B. Ansarinejad[8], M. Archipley[1,9], L. Balkenhol[10], K. Benabed[10], A. N. Bender[6,11], B. A. Benson[5,6,7], F. Bianchini[12,13,14], L. E. Bleem[6,11], F. R. Bouchet[10], L. Bryant[15], E. Camphuis[10], J. E. Carlstrom[6,7,11,15,16], T. W. Cecil[11], C. L. Chang[6,7,11], P. Chaubal[8], P. M. Chichura[6,16], T.-L. Chou[6,16], A. Coerver[2], T. M. Crawford[6,7], A. Cukierman[12,13,14], C. Daley[1], T. de Haan[17], K. R. Dibert[6,7], M. A. Dobbs[18,19], A. Doussot[10], D. Dutcher[3], W. Everett[20], C. Feng[21], K. R. Ferguson[22], K. Fichman[6,16], S. Galli[10], A. E. Gambrel[6], R. W. Gardner[15], F. Ge[23], N. Goeckner-Wald[12,13], R. Gualtieri[24], F. Guidi[10], N. W. Halverson[25,26], E. Hivon[10], G. P. Holder[21], W. L. Holzapfel[2], J. C. Hood[6], N. Huang[2], F. Kéruzoré[11], L. Knox[23], M. Korman[27], K. Kornoelje[6,7,11], C.-L. Kuo[12,13,14], A. T. Lee[2,28], K. Levy[8], A. E. Lowitz[6], C. Lu[21], A. Maniyar[12,13,14], F. Menanteau[1,9], M. Millea[2], J. Montgomery[18], Y. Moon[1], Y. Nakato[13], T. Natoli[6], G. I. Noble[29,30], V. Novosad[31], Y. Omori[6,7], S. Padin[6,32], Z. Pan[6,11,16], P. Paschos[15], K. A. Phadke[1,9], K. Prabhu[23], Z. Qu[1], W. Quan[6,16], M. Rahimi[8], A. Rahlin[5,6], C. L. Reichardt[8], C. Reuter[1], M. Rouble[18], J. E. Ruhl[27], E. Schiappucci[8], G. Smecher[33], J. A. Sobrin[5,6], A. A. Stark[34], J. Stephen[15], A. Suzuki[28], K. L. Thompson[12,13,14], B. Thorne[23], C. Trendafilova[9], C. Tucker[4], C. Umilta[21], J. D. Vieira[1,9,21], Y. Wan[1,9], G. Wang[11], N. Whitehorn[35], W. L. K. Wu[12,14], V. Yefremenko[11], M. R. Young[5,6], and J. A. Zebrowski[5,6,7]

[1] Department of Astronomy, University of Illinois Urbana–Champaign, 1002 West Green Street, Urbana, IL 61801, USA; ctandoi2@illinois.edu
[2] Department of Physics, University of California, Berkeley, Berkeley, CA 94720, USA
[3] Joseph Henry Laboratories of Physics, Jadwin Hall, Princeton University, Princeton, NJ 08544, USA
[4] School of Physics and Astronomy, Cardiff University, Cardiff CF24 3YB, UK
[5] Fermi National Accelerator Laboratory, MS209, P.O. Box 500, Batavia, IL 60510, USA
[6] Kavli Institute for Cosmological Physics, University of Chicago, 5640 South Ellis Avenue, Chicago, IL 60637, USA
[7] Department of Astronomy and Astrophysics, University of Chicago, 5640 South Ellis Avenue, Chicago, IL 60637, USA
[8] School of Physics, University of Melbourne, Parkville, VIC 3010, Australia
[9] Center for AstroPhysical Surveys, National Center for Supercomputing Applications, Urbana, IL 61801, USA
[10] Sorbonne Université, CNRS, UMR 7095, Institut d'Astrophysique de Paris, 98 bis bd Arago, 75014 Paris, France
[11] High-Energy Physics Division, Argonne National Laboratory, 9700 South Cass Avenue, Lemont, IL 60439, USA
[12] Kavli Institute for Particle Astrophysics and Cosmology, Stanford University, 452 Lomita Mall, Stanford, CA 94305, USA
[13] Department of Physics, Stanford University, 382 Via Pueblo Mall, Stanford, CA 94305, USA
[14] SLAC National Accelerator Laboratory, 2575 Sand Hill Road, Menlo Park, CA 94025, USA
[15] Enrico Fermi Institute, University of Chicago, 5640 South Ellis Avenue, Chicago, IL 60637, USA
[16] Department of Physics, University of Chicago, 5640 South Ellis Avenue, Chicago, IL 60637, USA
[17] High Energy Accelerator Research Organization (KEK), Tsukuba, Ibaraki 305-0801, Japan
[18] Department of Physics and McGill Space Institute, McGill University, 3600 Rue University, Montreal, QC H3A 2T8, Canada
[19] Canadian Institute for Advanced Research, CIFAR Program in Gravity and the Extreme Universe, Toronto, ON M5G 1Z8, Canada
[20] Department of Astrophysical and Planetary Sciences, University of Colorado, Boulder, CO 80309, USA
[21] Department of Physics, University of Illinois Urbana–Champaign, 1110 West Green Street, Urbana, IL 61801, USA
[22] Department of Physics and Astronomy, University of California, Los Angeles, Los Angeles, CA 90095, USA
[23] Department of Physics & Astronomy, University of California, Davis, One Shields Avenue, Davis, CA 95616, USA
[24] Department of Physics and Astronomy, Northwestern University, 633 Clark Street, Evanston, IL 60208, USA
[25] CASA, Department of Astrophysical and Planetary Sciences, University of Colorado, Boulder, CO 80309, USA
[26] Department of Physics, University of Colorado, Boulder, CO 80309, USA
[27] Department of Physics, Case Western Reserve University, Cleveland, OH 44106, USA
[28] Physics Division, Lawrence Berkeley National Laboratory, Berkeley, CA 94720, USA
[29] Dunlap Institute for Astronomy & Astrophysics, University of Toronto, 50 St George Street, Toronto, ON M5S 3H4, Canada
[30] David A. Dunlap Department of Astronomy & Astrophysics, University of Toronto, 50 St George Street, Toronto, ON M5S 3H4, Canada
[31] Materials Sciences Division, Argonne National Laboratory, 9700 South Cass Avenue, Lemont, IL 60439, USA
[32] California Institute of Technology, 1200 East California Boulevard, Pasadena, CA 91125, USA
[33] Three-Speed Logic, Inc., Victoria, BC V8S 3Z5, Canada
[34] Center for Astrophysics | Harvard & Smithsonian, 60 Garden Street, Cambridge, MA 02138, USA
[35] Department of Physics and Astronomy, Michigan State University, East Lansing, MI 48824, USA

## Abstract

We present a flare star catalog from 4 yr of nontargeted millimeter-wave survey data from the South Pole Telescope (SPT). The data were taken with the SPT-3G camera and cover a 1500 deg$^2$ region of the sky from $20^{h}40^{m}0^{s}$ to $3^{h}20^{m}0^{s}$ in right ascension and from $-42^{\circ}$ to $-70^{\circ}$ in declination. This region was observed on a nearly daily cadence from 2019 to 2022 and chosen to avoid the plane of the galaxy. A short-duration transient search of this survey yields 111 flaring events from 66 stars, increasing the number of both flaring events and detected flare stars by an order of magnitude from the previous SPT-3G data release. We provide cross-matching to Gaia DR3, as well as matches to X-ray point sources found in the second ROSAT all-sky survey. We have detected flaring stars across the main sequence, from early-type A stars to M dwarfs, as well as a large population of evolved stars. These stars are mostly nearby, spanning 10–1000 pc in distance. Most of the flare spectral indices are constant or gently rising as a function of frequency at 95/150/220 GHz. The timescale of these events can range from minutes to hours, and the peak $\nu L_\nu$ luminosities range from $10^{27}$ to $10^{31}$ erg s$^{-1}$ in the SPT-3G frequency bands.

## 1. Introduction

Flaring activity in stars can be found across a wide range of spectral types (Balona 2015) and is thought to result from a release of magnetic energy. The Sun is also known to flare and has been extensively studied from radio wavelengths to gamma rays (Fletcher et al. 2011; Benz 2017), including the millimeter wavelengths (Silva et al. 1996) studied in this paper. Solar flares have been used to serve as a framework for our understanding of flaring activities on other stars (Pettersen 1989). While there are similarities in processes and the physics between many solar and stellar flares (Aschwanden et al. 2008), the utility of these comparisons reaches a limit, as the Sun is atypical when looking at larger stellar populations. For example, even when considering other "Sun-like" stars (stars with similarities such as spectral type, age, rotation period, and position along the main sequence), the Sun has been shown to exhibit substantially lower photometric variability (Reinhold et al. 2020). Typical solar flares have optical energies of $\sim 10^{29}$ erg with a maximum of $\sim 10^{31}$ erg in X-ray. On the extreme side, flares as strong as $\sim 10^{38}$ erg at optical wavelengths have been observed on "Sun-like" stars (Schaefer et al. 2000).

The generally accepted model of stellar flares (Brown 1971; Hudson 1972) involves a sudden release of energy triggered by reconnection of magnetic field lines, resulting in acceleration of charged particles that emit radiation across the electromagnetic spectrum (Allred et al. 2015; Benz 2017). Different wavelengths track different physics of the flare: charged particles are accelerated from this area of reconnection in an impulsive event, seen simultaneously as gyrosynchrotron in radio wavelengths and bremsstrahlung in hard X-rays (HXR, $\gtrsim 10$ keV; Bastian et al. 1998; Massi et al. 2002; Benz & Güdel 2010). The particles travel along the field lines, eventually depositing their energy lower in the stellar atmosphere. This energy heats and evaporates plasma back into the upper atmosphere and can be seen in optical/ultraviolet (UV; Henry & Newsom 1996; Krucker et al. 2015). As this plasma reaches the corona, the emission transitions from a sudden and fast rise in energy released to a gradual rise that can be seen in soft X-rays (SXR, $\lesssim 10$ keV) as cooling starts (Güdel 2004).

External interactions between stars can also produce flaring events. RS Canum Venaticorum (RS CVn) variable stars and other binary systems have shown flares across the electromagnetic spectrum, which can be due to interactions in the magnetic field shared between components of the system (Catalano 1986; Massi et al. 2002). Protostars and young, pre-main-sequence stars (e.g., T Tauri) that still have an accretion disk can exhibit flares owing to magnetic field interactions between the star and the disk itself (Feigelson & Montmerle 1999).

At radio wavelengths, solar flares typically show a peak at 1–10 GHz, which is thought to correspond to mildly relativistic synchrotron radiation as the source transitions from optically thick to optically thin, with the optically thin part of the spectrum extending into millimeter wavelengths (Bastian et al. 1998). Peak flux spectra of solar flares show flux at 5 GHz to be over an order of magnitude larger than flux at 30 GHz (Kundu & Vlahos 1982), with more recent observations exhibiting a similar falling spectrum between flux at 3.4–17 GHz and 86 GHz (Raulin et al. 1999) and even out to 345 GHz (Lüthi et al. 2004). Rising spectra at submillimeter wavelengths have been occasionally observed in solar flares (Zhou et al. 2011), with radiative processes such as bremsstrahlung and gyrosynchrotron requiring extreme parameters as possible explanations (Krucker et al. 2013).

Stellar flare astronomy has historically involved targeted observations of individual stars (e.g., Webber et al. 1973; Large et al. 1989). Recent years have seen massive growth in the field of time-domain astronomy, with exoplanet-detecting experiments Kepler (Koch et al. 2010), and TESS (Ricker et al. 2014) providing observing cadences on the order of minutes. These high-cadence data are invaluable for monitoring the variability of stars and have resulted in enormous catalogs of stellar flares at optical wavelengths (Davenport 2016; Günther et al. 2020). Transient surveys have also been carried out at infrared (Davenport et al. 2012), UV (Loyd et al. 2018; Brasseur et al. 2019), and X-ray (Pye et al. 2015; Getman & Feigelson 2021).

At longer wavelengths, radio observations of stellar flares have been mainly composed of targeted observations (Smith et al. 2005; Loyd et al. 2018) and recent archival surveys (Levinson et al. 2002). In the future, radio wavelength surveys such as Square Kilometre Array (SKA) pathfinders, and eventually SKA, will monitor the southern sky for radio transients (Murphy et al. 2013).

While observations of stellar flares at or near millimeter wavelengths have been made in the past (Beasley & Bastian 1998; Brown & Brown 2006; Massi et al. 2006; Umemoto et al. 2009; MacGregor et al. 2018; Mairs et al. 2019; MacGregor et al. 2020, 2021; Burton et al. 2022), there have not been any nontargeted transient surveys at these wavelengths until those conducted with the South Pole Telescope (SPT Whitehorn et al. 2016; Guns et al. 2021), which yielded 13 stellar flares from eight stars, and the Atacama Cosmology Telescope (ACT; Naess et al. 2021; Li et al. 2023), which yielded a total of 17 stellar flares from 14 stars.

Stellar flares detected by SPT and ACT have been in the range of $\nu L_\nu \sim 10^{26}$–$10^{30}$ erg s$^{-1}$ at 150 GHz, with durations on the scale of minutes to hours and spectral indices typically in the range of −1 to +2. These flares are orders of magnitude brighter and longer than recent Atacama Large Millimeter/submillimeter Array (ALMA) observations of millimeter stellar flares provided in Table 1 of Burton et al. (2022): 10 flares from the stars au Mic, Proxima Centauri, and Epsilon Eridani, which have luminosities of $\nu L_\nu \sim 10^{24}$–$10^{26}$ erg s$^{-1}$ with durations on the scale of seconds. Burton et al. (2022) also report a wider range of spectral indices for these ALMA flares, ranging from roughly −3 to 7. Solar flares as a comparison are even weaker: one of the brightest solar flares ever recorded—an X28-class flare detected on 2003 November 4—had a 212 GHz luminosity of $\nu L_\nu \sim 10^{23}$ erg s$^{-1}$ (Zhou et al. 2011).

In this paper we perform a nontargeted search using data from the 2019–2022 observing seasons of the SPT-3G survey and present a catalog of 111 flares from 66 stars, an increase of almost a factor of 10 over the previous transient analysis of the 2020 observing season presented in Guns et al. (2021). Providing such a large sample of these very strong millimeter flares can help improve our understanding of the physics during the impulsive particle acceleration phase of stellar flares. M dwarfs are of particular interest from the standpoint of exoplanet habitability (Shields et al. 2016), and millimeter emission as a proxy for far-UV emission during stellar flares may help constrain the radiation environment around exoplanets (MacGregor et al. 2021). Additionally, characterizing this population of flaring stars will be useful for future millimeter wavelength transient surveys, such as CMB-S4 (Abazajian et al. 2019) and the Simons Observatory (SO; Ade et al. 2019).

The format of the paper is as follows: Mapmaking and filtering along with transient detection methods are described in Section 2. The flare star catalog, full 4 yr single-observation light curves at 95 and 150 GHz for each associated source, and single-scan light curves at 95, 150, and 220 GHz of each flaring event are provided in Section 3, as well as characterizations of the stellar populations and analysis of flaring properties. A conclusion including directions for further study is described in Section 4.

## 2. Methods

### 2.1. Instrument and Survey

The SPT is a 10 m telescope located at Amundsen−Scott South Pole Station, Antarctica (Carlstrom et al. 2011). The SPT-3G receiver is the third camera to be mounted on the SPT and has been operational since 2017. The SPT-3G focal plane contains 2690 dual-polarization, trichroic pixels sensitive in three frequency bands centered at 95, 150, and 220 GHz. Each pixel contains six bolometric detectors, for a total of ∼16,000 detectors read out using a frequency-domain multiplexing system. The large 10 m primary on the SPT allows for high-resolution observations, with beam sizes on the order of 1′ FWHM (Sobrin et al. 2022).

The data used in this work are from 4 yr of the SPT-3G winter survey, which comprises a 1500 deg$^2$ patch of the southern sky and is observed annually from late March through November with an average cadence of 12 hr. The SPT-3G winter field spans a rectangle in sky coordinates from $20^{\rm h}40^{\rm m}0^{\rm s}$ to $3^{\rm h}20^{\rm m}0^{\rm s}$ in R.A. and from −70° to −42° in decl. The field is split horizontally into four equal height subfields. A single SPT-3G observation consists of a series of scans across a given subfield that last approximately 2 hr. During an observation, the SPT rasters back and forth across the sky at constant elevation and scan speed (constant on the bearing, i.e., in azimuth angle per second) before taking a 12.′5 step in elevation and repeating the process, until the entire subfield has been observed. A full description of the design, operation, and performance of the SPT-3G camera can be found in Sobrin et al. (2022).

We use a set of SPT-3G winter survey maps that have been optimized for source finding. Each single-frequency, single-observation subfield map is made by filtering the detector time-ordered data and using individual detector pointing to bin the data into a pixelated 2D sky map with 0.′25 pixels in a Lambert azimuthal equal-area projection. The map filtering is described in detail in Dutcher et al. (2021), and we follow the same procedure but with different high-pass and low-pass cutoff values for the time-domain filter, which are set to filter out modes with spatial scales below angular multipoles of $\ell = 500$ and above $\ell = 20{,}000$. Bright point sources (defined as having average flux ⩾ 50 mJy at 150 GHz) are linearly interpolated over before the time-domain filter in order to suppress filtering artifacts, and the interpolated samples are removed from the search.

Similar to Guns et al. (2021), we subtract a 3 yr average map from each individual observation in order to remove the cosmic microwave background (CMB), galaxy clusters, static point sources, and other static backgrounds. The background-subtracted maps thus contain a combination of red atmospheric noise that dominates at large angular scales (typically $\ell < 2000$, but up to $\ell = 3000$ for adverse weather conditions), detector white noise, and time-varying point-source signals. We apply a map-space pixel mask to remove a 5′-radius disk around all 2465 known point sources with 3 yr average fluxes of 5 mJy or higher at 150 GHz from the search area. In order to remove the atmospheric noise, we apply an isotropic 2D high-pass filter as a weighted convolution in map space (using inverse-variance weights) and filter the resulting white-noise-dominated maps with a Gaussian beam template to maximize sensitivity to point sources. The beam template is constructed as a Gaussian fit to the central lobe of the SPT-3G beams, with widths $\sigma_{\rm beam} = 0.'67$, $0.'48$, and $0.'45$ at 95, 150, and 220 GHz, respectively. The isotropic high-pass filter kernel is an annulus with an inner radius of $3\sigma_{\rm beam}$ and an outer radius of 5′ at 95 GHz and 3′ at 150 and 220 GHz. More aggressive filtering is required at higher frequencies because of the rising atmospheric noise spectrum. Unlike static SPT source analyses, which use multiyear averaged maps and carry out all map-space filtering operations in the Fourier domain, we choose to filter in real space because the pixel weights in our single-observation maps are nonuniform (varying by as much as 20% across the map), which is properly accounted for with a weighted real-space convolution filter. The time-domain filter imprints additional large-scale structure onto the beam, adding degree-scale filtering artifacts along lines of constant declination. As the computing time for a real-space convolution increases linearly with the kernel area, we choose not to account for this large-scale structure. Compared to the full-size filter kernel, this choice sacrifices ∼3.5% in optimality in order to speed up the filtering operations by an order of magnitude.

We summarize the pixel noise for each single-observation map by taking the median noise across all the pixels in the map. The results are shown in Figure 1. Across all observations in all subfields, we achieve median noise values of 7.8 mJy rms at 95 GHz and 8.0 mJy rms at 150 GHz. The median noise at 220 GHz is much higher at ∼27 mJy rms, with this noise discrepancy described in more detail in Sobrin et al. (2022). As a result, we do not include the 220 GHz maps in our transient-finding method and only use them to measure fluxes of sources detected in the other two observing bands. The constant scan speed in azimuth results in a slower on-sky scan speed and longer observation time per map area with increasing elevation (decreasing decl.). This, together with higher atmospheric loading (and corresponding higher detector noise), results in higher map noise at lower elevations. To achieve uniform noise across the whole winter survey field on long timescales, the lower elevation fields are observed more frequently. Although the cadence averages out to one full field observation every 12 hr, the actual cadence consists of clusters of two or three observations of the same subfield followed by either switching to a different elevation or a period of downtime to cycle the cryogenic fridge and observe calibration sources. As a consequence, the actual time between subsequent observations of the same source can range from 2 hr to approximately 36 hr.

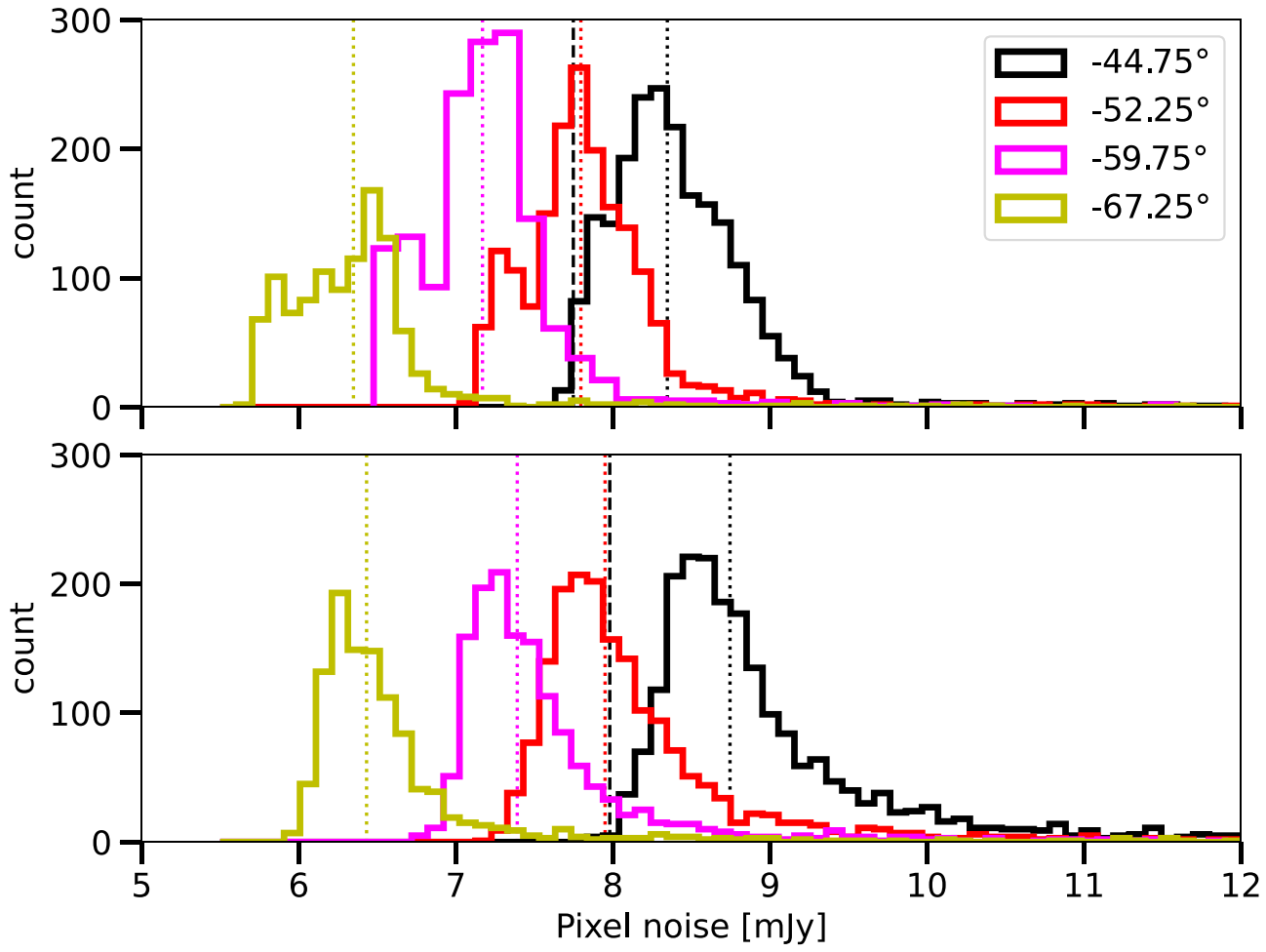

**Figure 1.** Histograms of noise values for every observation processed in the archival transient pipeline, grouped by subfield. Histograms are colored by subfield and referred to by the center decl. of that subfield. The noise value for each observation is the median pixel noise across all the pixels in the corresponding sky map. The top and bottom panels show the noise values for the 95 and 150 GHz maps, respectively. Dotted colored lines mark the median noise across all observations within a given subfield, while the dashed black line shows the overall median across all observations.

The nonuniform observing cadence and uneven subfield coverage between reobservations of the same subfield make it necessary to use simulations to calculate the observing efficiency for bright transients of specific durations. By injecting simple boxcar models of flares brighter than the single-observation detection threshold, we find that our instrument and observing strategy detect approximately 80% of 24 hr flares, 20% of 4 hr flares, and less than 5% of flares in our field lasting 30 minutes or less.

### 2.2. Transient Finding

The transient-finding method is described in detail in Guns et al. (2021), and we provide a brief overview here, as well as noting any pertinent changes. For each unmasked location on the pixelated sky, we construct a series of two-band light curves spanning consecutive 12-day intervals and using only 95 and 150 GHz data. The 12-day interval covers short-duration signals of interest—millimeter stellar flares from previous and ongoing analyses (i.e., Guns et al. 2021), as well as predicted short-duration extragalactic millimeter transients (Eftekhari et al. 2022). We apply a preliminary cut on the data, selecting all light curves that show a $>3\sigma$ excursion in both bands at the same point in time. To each remaining light curve we fit a four-parameter Gaussian flare model with two independent flux parameters $S_{95}$ and $S_{150}$ at 95 and 150 GHz, a peak time $t_0$, and an event FWHM $w$. The test statistic (TS) is constructed by maximizing the likelihood ratio of the best-fit flare model to the null model (zero amplitude in both bands), with an additional logarithmic penalty term. This penalty term corrects for a likelihood bias toward short flare widths: since there are many more uncorrelated starting times for short flare widths than for long ones, the effective volume of searchable parameter space grows as the flare width decreases. The logarithmic penalty term effectively imposes a flat prior on the flare duration in order to correct for this effect (Braun et al. 2010).

The TS is thus defined as

$$\mathrm{TS} = 2\ln\mathcal{L}(\hat{S}_{95}, \hat{S}_{150}, \hat{t}_0, \hat{w}) - 2\ln\mathcal{L}(0) + 2\ln\left(\frac{\hat{w}}{12\ \mathrm{days}}\right), \tag{1}$$

where quantities with hats denote best-fit parameters as found by a nonlinear optimizer using the Nelder–Mead simplex method, as implemented in the GNU Scientific Library (Galassi et al. 2009).

Focusing our search on short-duration events, we impose a significance cut of $\mathrm{TS} > 45$ and a best-fit FWHM duration of $<10$ days. The TS cutoff is set by demanding a negligible number of false positives from background noise. At $\mathrm{TS} > 45$ we expect to see $\sim 1$ noise-induced event per year (S. Guns et al. 2024, in preparation). In combination with the Gaia association method of Section 3.1.2, this should lead to $<0.1$ noise-induced flaring events with a confident Gaia cross-match per year. In the short-duration regime, the main systematic background for SPT transients are weather balloons that are launched daily by the South Pole Meteorological Office (Met). Through a data-sharing agreement established in 2022, SPT now receives Met GPS data soon after each balloon launch and cuts any telescope scans that intersect with a weather balloon trajectory, vastly reducing the number of detected balloons in the SPT-3G transient detection pipeline at the cost of a negligible reduction in data volume. The balloon avoidance system is not infallible, however, and fails when the tracking data arrive too late to be integrated into the SPT-3G data processing pipeline, or when the balloon ruptures at high altitude and debris that is not tracked by the balloon's GPS module enters the SPT-3G field of view. Remaining balloon detections number approximately one per month and are identified by their observational signature: due to their fast movement, they are detected with a short duration in single telescope scans (lasting 2 s or less), with high brightness ($>1$ Jy), with thermal or steeper-than-thermal spectra, and within 2 hr of a logged balloon launch.

After removing events with best-fit durations $>10$ days and $\mathrm{TS} < 45$, we are left with a data set of candidate events that we inspect by eye. We manually analyze the light curves and thumbnails to reject duplicate observations of the same event, filtering artifacts from bright nearby transients, untracked balloons, and detector glitches (large map streaks in the scan direction). Events with any ambiguity are excluded from further analysis. After these cuts, we keep 111 short-duration ($<10$-day) events with robust morphologies that we use to cross-match to the Gaia Data Release 3 (DR3) catalog (Gaia

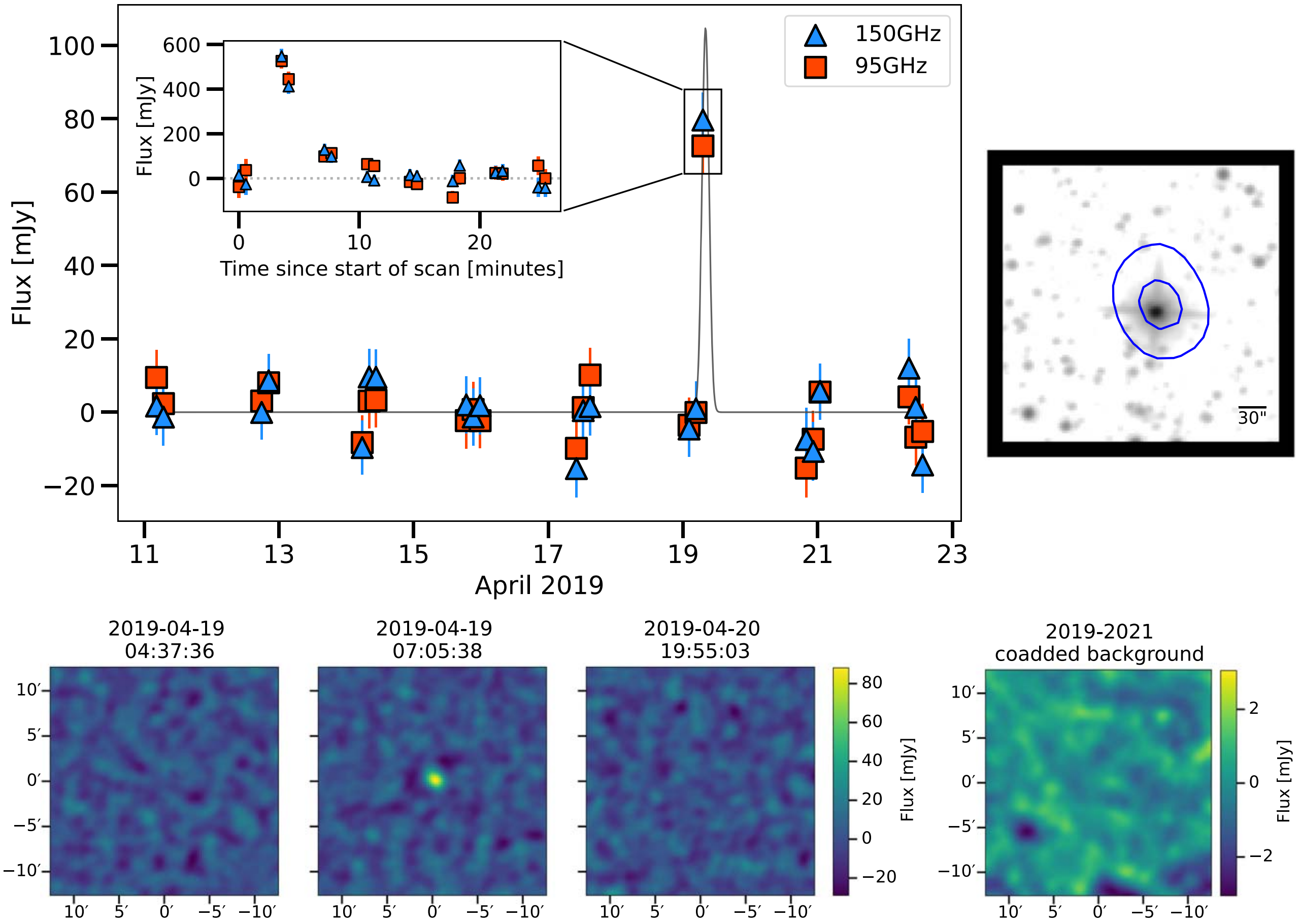

**Figure 2.** Example light curves and thumbnails for the SPT event on MJD 58592.3. Top left: light curves for the highest TS pixel on the triggered event. The main plot shows the peak single-observation flux, while the inset shows the single-scan flux within the peak single observation. Top right: unWISE 3.4 $\mu$m W1 grayscale image (Lang 2014) with blue SPT 150 GHz flux density contours in 5$\sigma$ steps from the peak signal. Bottom: 150 GHz thumbnails for the single-observation data points (immediately before, during, and immediately after the triggered event), and the 3 yr average map that has been subtracted to create these thumbnails. These thumbnails show that this flaring star is undetectable in single SPT-3G observations at its quiescent level.

Collaboration et al. 2023). For these 111 events, we have performed tests using different flare models for transient detection to compare viability. Using a standard flare model (Mendoza et al. 2022) returned slightly lower TS values on average when compared to our Gaussian fit, while using a basic exponential decay model resulted in a slight scatter in TS values when compared to the Gaussian fit. A combination of fit models could result in a higher completeness of detected flares in our data.

Figure 2 shows an example event detected by the transient pipeline. The main light-curve plot shows the data that are input into the light-curve fitter and the resulting Gaussian fit, with a TS of 194.8. Each data point represents the average flux over a 2 hr SPT subfield observation, in which any given point on the sky is observed for about eight consecutive left/right scans (16 scans total) over the course of 20 minutes. We construct post-detection light curves of the single-scan fluxes (each scan representing one pass of the focal plane, or roughly 2 s of integrated data) that make up the peak observation of the flare, shown as an inset to the main light curve. Figure 3 shows thumbnails of the single scans for this event. Thumbnails of scans 1, 2, 19, and 20 are discarded because the sky location of interest is not observed by the focal plane. Scans at the beginning and end (such as scans 3 and 4) are noisier, due to only part of the focal plane seeing the sky location of interest.

In order to localize events at finer scales than the 15″ pixel grid, we determine an event centroid location from the TS map associated with each flare. The TS map is constructed by fitting the light-curve model to each pixel in a $5 \times 5$ pixel ($1\farcm25 \times 1\farcm25$) box centered on the event. The TS map combines all the available information from every time point and both frequency bands (under the assumption of a Gaussian flare model). For each event, we fit a cubic spline to the TS map and choose the spline maximum as the event location.

The uncertainties in the SPT R.A. and decl., $\sigma_{\rm R.A.}$ and $\sigma_{\rm decl.}$, come from a combination of systematic pointing uncertainty and statistical uncertainty in localizing a source given the SPT beam and noise:

$$\sigma^2_{\rm R.A.} = 3.73''^2 + \left(\frac{50''}{\sqrt{\rm TS}}\right)^2, \qquad (2)$$

$$\sigma^2_{\rm decl.} = 4.35''^2 + \left(\frac{49''}{\sqrt{\rm TS}}\right)^2. \qquad (3)$$

The first term is the systematic contribution to the error budget and comes from uncertainties in the pointing of the physical instrument, whereas the second term is the statistical contribution and is sourced by the noise in the observations. We estimate the first term by measuring the positions of 1 yr of daily observations of bright point sources and comparing to a reference catalog. The second term is calculated from injecting $N = 10{,}000$ simulated sources of various amplitudes at known locations into the SPT field and recovering the measured position. In a small fraction (<1%) of cases, heavy winds and other inclement weather lead to systematic pointing errors that are much larger than the mean estimate in Equations (2) and (3). In those cases, we estimate the systematic pointing error from the bright point-source positions as measured in that observation only.

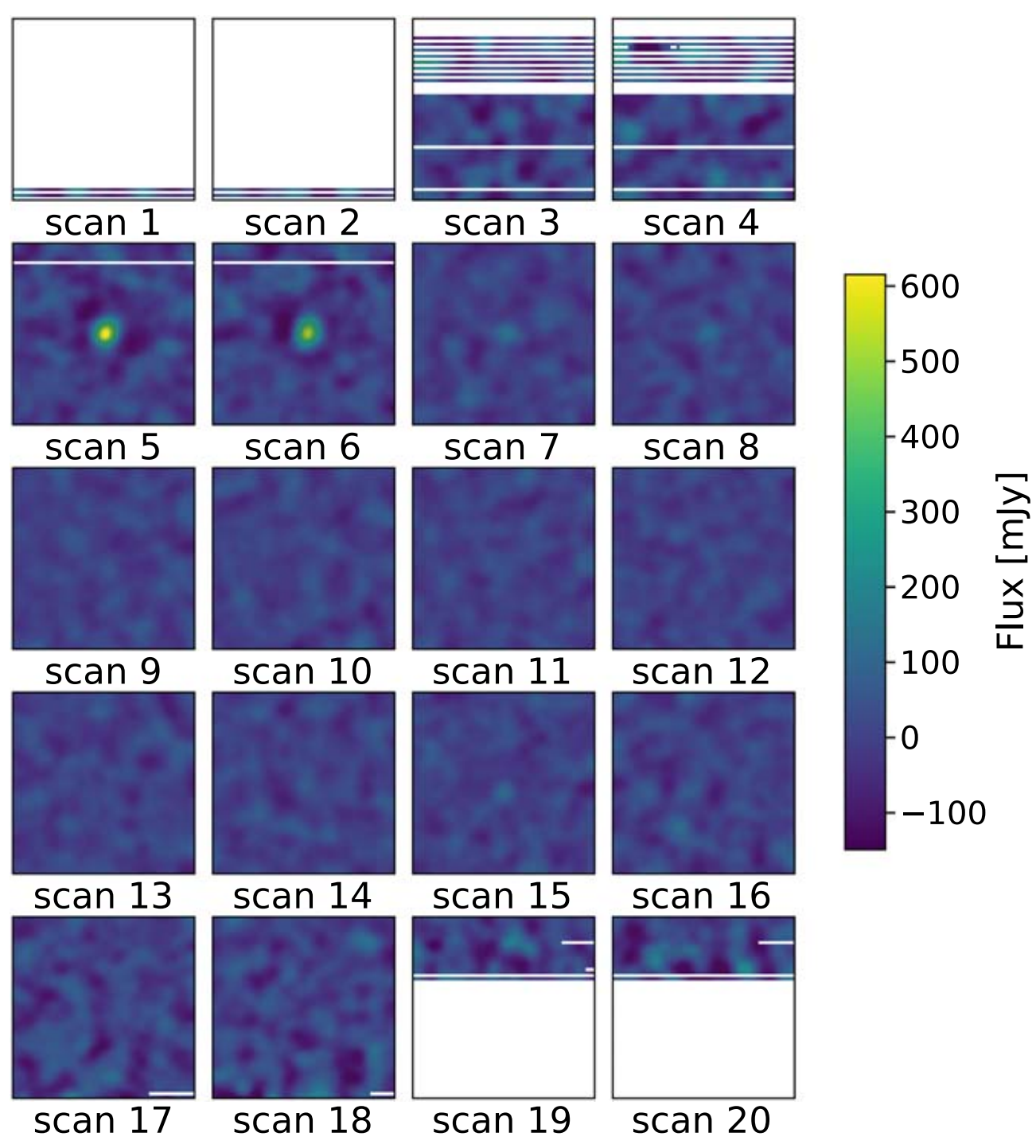

**Figure 3.** $15' \times 15'$ single-scan 150 GHz thumbnails centered on the highest TS pixel from the event in Figure 2. The source is only clearly visible in scans 5 and 6 and has returned to near noise flux values in scans 7 and 8.

## 3. Results

### 3.1. Stellar Association

#### 3.1.1. Candidate Sources

To find candidate stars for our events, we only consider Gaia sources within our observing field that have a maximum $G$ magnitude of 20 and exclude sources that are found in the qso_candidates or galaxy_candidates tables in Gaia DR3 to eliminate known extragalactic sources. This leaves about 7.3 million stars as possible matches for SPT-3G transients. X-ray associations with sources from the second ROSAT all-sky survey (2RXS) are provided by the main table of the ROSAT stellar content catalog (Freund et al. 2022), matching only on Gaia DR3 source ID. For Gaia stars without a match in the stellar content catalog, we check the full 2RXS point-source catalog (Boller et al. 2016) for any sources within 1′ and have separately distinguished them.

#### 3.1.2. Cross-matching Algorithm

To associate flaring events with sources, the cross-matching algorithm we use compares the likelihood of an event originating from a candidate source to the likelihood of a random event having a chance association, using a 1° radius centered on each candidate to determine a local source density in the Gaia catalog. This results in the figure of merit in association $\Lambda$:

$$\Lambda = \ln\left(\mathcal{B}\frac{N_{\text{total}}}{N(>S)}\right), \tag{4}$$

where $N_{\text{total}}$ is the number of sources within the local area and $N(>S)$ is the number of sources within the local area that are brighter in Gaia magnitude than the candidate source. $\mathcal{B}$ considers the uncertainties of the SPT beam and the angular separation:

$$\mathcal{B} = \exp\left[-\frac{1}{2}\left(\left(\frac{\Delta_{\text{R.A.}}}{\sigma_{\text{R.A.}}}\right)^2 + \left(\frac{\Delta_{\text{decl.}}}{\sigma_{\text{decl.}}}\right)^2\right)\right]. \tag{5}$$

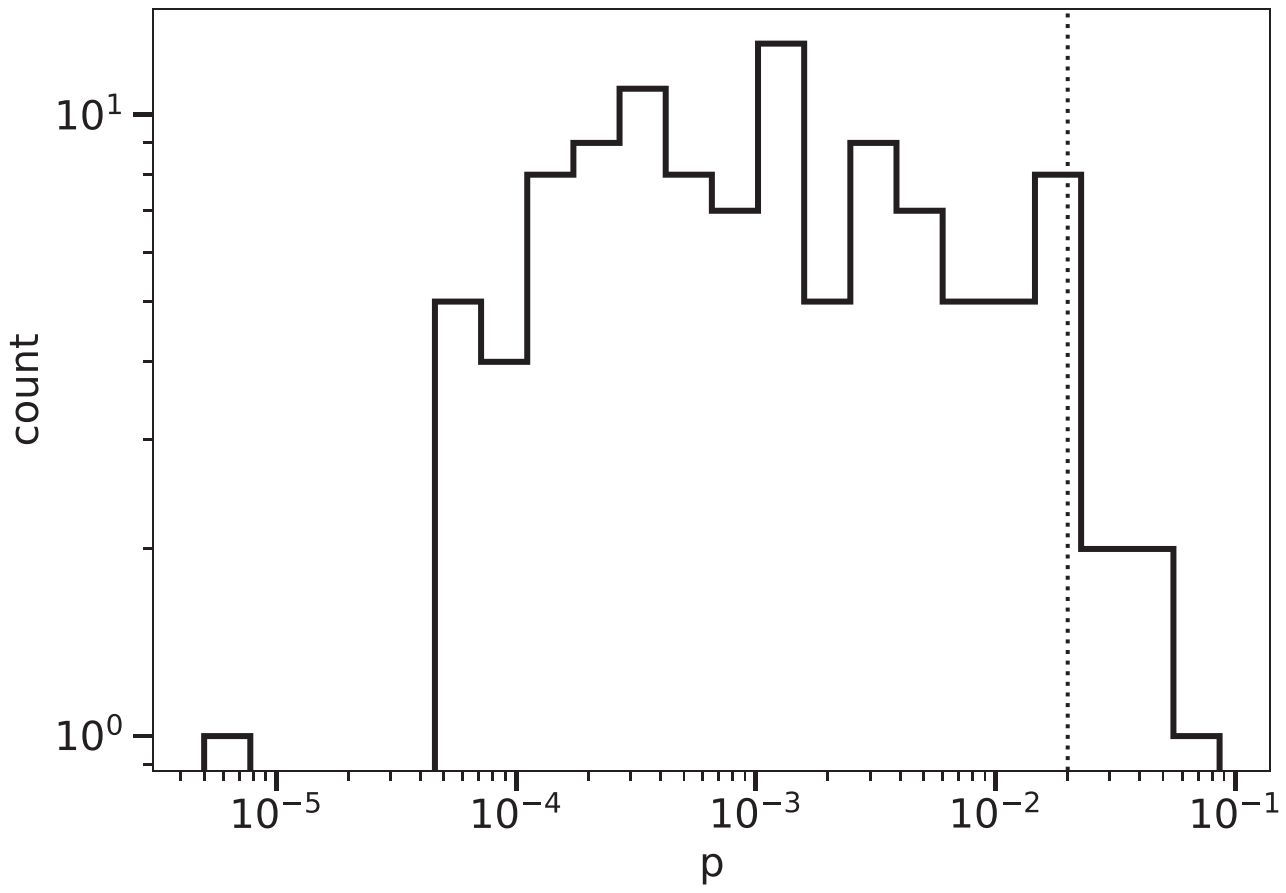

**Figure 4.** The distribution of $p$-values from matching SPT flaring events to stars in the Gaia DR3 catalog. The five events at $p > 0.02$ have been investigated and are discussed in Section 3.8.3.

$\Delta_{\text{R.A.}}$ and $\Delta_{\text{decl.}}$ are the angular separations in R.A. and decl. between the transient centroid location and candidate source in the Gaia catalog, while $\sigma_{\text{R.A.}}$ and $\sigma_{\text{decl.}}$ are defined in Equations (2) and (3).

We create a probability distribution function for random matches by sampling 400,000 random points in the SPT-3G field and matching them to Gaia using this algorithm. We then construct a cumulative distribution function (CDF) based on the distribution of $\Lambda$ values for this random sample, and then we create a one-to-one mapping of $\Lambda$ to $p$-values in the form of $p(\Lambda) = 1 - \text{CDF}(\Lambda)$. We interpolate this mapping to find the $p$-value corresponding to the observed $\Lambda$ for each event, telling us the chance of a random event having that same association value or higher. The distribution of $p$-values for our events is shown in Figure 4.

### 3.2. SPT-3G Flare Star Catalog

As described in Section 2.2, this flare star catalog is a result of searching only for short-duration transient events in SPT-3G archival data, using a cutoff of 10 days on the best-fit FWHM duration. With our current analysis we have recovered 12 of the 13 stellar flaring events in Guns et al. (2021), and we include them in the flare star catalog. The missing flaring event, previously referred to as event 6(a) by the star CC Eri, was a weak flare detected at 29/21 mJy in the 95/150 GHz bands with respective luminosities of $4.4 \times 10^{26}/5.0 \times 10^{26}$ erg s$^{-1}$ and a TS of 29.5, below the current detected threshold of 45.

**Table 1**
Columns and Descriptions for the SPT-3G Flare Star Catalog

| Column | Label | Units | Description |
|---|---|---|---|
| 1 | spt_id | | IAU-approved SPT source name, corresponding to the SPT-detected location. |
| 2 | mjd | days | The approximate time in MJD corresponding to the beginning of the observation the detection took place in. |
| 3 | ts | | The test statistic of the event, used to flag a possible flaring event occurring during that observation. Minimum values are 45, with a higher value having a more significant detection. |
| 4, 5 | ra<br>dec | deg | The R.A. and decl. (J2000) of the SPT event. |
| 6–11 | 95_flux<br>95_flux_err<br>150_flux<br>150_flux_err<br>220_flux<br>220_flux_err | mJy | Flux values and their errors for the three SPT bands. |
| 12–15 | spectral_index_95_150<br>spectral_index_95_150_err<br>spectral_index_150_220<br>spectral_index_150_220_err | | Two-band spectral index values and their errors for the 95/150 and 150/220 SPT bands. |
| 16 | dr3_source_id | | Gaia DR3 unique source identifier. |
| 17 | p | | The probability of a random event cross-matching to this source. |
| 18, 19 | source_ra<br>source_dec | deg | The R.A. and decl. (J2000) of the Gaia DR3 source. |
| 20–22 | phot_g_mean_mag<br>phot_bp_mean_mag<br>phot_rp_mean_mag | | Magnitudes of the three Gaia passbands. |
| 23, 24 | parallax<br>parallax_err | mas | Parallax and error of the Gaia source. |
| 25 | binary | | Boolean flag for binary candidacy. |
| 26 | 2rxs_id | | 2RXS identifier. |
| 27 | 2rxs_ref | | Flag for source association: "Freund" for an established association,[a] or "DR3/2RXS Proximity" for nearby sources. |
| 28 | 2rxs_flux | mW m$^{-2}$ | 2RXS flux. |

**Note.**
[a] Freund et al. (2022).

(This table is available in its entirety in machine-readable form in the online article.)

The other flares from CC Eri were successfully recovered. The previous analysis in Guns et al. (2021) had a detection threshold of TS = 100. Any location in the sky exhibiting a flare with a TS $\geqslant$ 100 was then manually scrutinized over the entire observing season to see whether there were smaller flares below the threshold that were not detected. This method was how event 6(a), at TS = 29.5, was found originally. We do not search for below-threshold flares in the current analysis and only show events above the new TS = 45 threshold. The two long-duration events from Guns et al. (2021) are longer than the 10-day best-fit FWHM cut and thus were not recovered.

Table 1 shows the full column names and descriptions for the flare star catalog. A full machine-readable table can be found in the online Journal and at the project webpage,[36] with associated light-curve data provided on Zenodo: doi:10.5281/zenodo.11372033.

After matching with the external catalogs described in Section 3.1.1, the full 4 yr SPT flare star catalog has associated all 111 flaring events with 66 unique stars. We find that 17 stars have flared multiple times and are responsible for 62 events, leaving 49 stars with single events associated with them. Distances for stars are derived from Gaia DR3 parallax when available. Five stars, associated with six events, have no reported parallax from Gaia. Table 2 shows alternate sources of parallax, or distance if parallax is not available.

[36] https://pole.uchicago.edu/public/data/tandoi24/

**Table 2**
Alternate Sources for Parallax or Distance for Stars That Have Missing Parallax in Gaia, with Rounded Values

| Gaia DR3 ID | Parallax (mas) | Distance (pc) | Source |
|---|---|---|---|
| 4914632365579875200 | 21 | … | a |
| 4726387007013852672 | 59 ± 1 | … | b |
| 6580179279486143744 | … | 44 ± 2 | c |
| 4850405974392619648 | … | 44 ± 1 | d |
| 4942856382389905408 | … | 19 ± 4 | e |

**Notes.**
[a] Riedel et al. (2014).
[b] van Leeuwen (2007).
[c] Bell et al. (2015).
[d] Stassun et al. (2019).
[e] Finch et al. (2014).

Figure 5 shows the population of SPT-detected flare stars with Gaia parallax in a color–magnitude diagram (CMD) and a magnitude versus distance plot. Different marker shapes denote our classifications of spectral types, with "Evolved" containing giants and subgiants, "Upper main sequence" containing early-type and Sun-like stars, and "Lower main sequence" mainly composed of M dwarfs. These classifications are based purely on position in the CMD. We note that these classifications may be inaccurate, due to the possibility of misrepresenting a binary system, which we discuss in Section 3.8. In addition, the two main-sequence outlier stars that we show in Section 3.8.3 are likely misidentifications and are not used in the analysis.

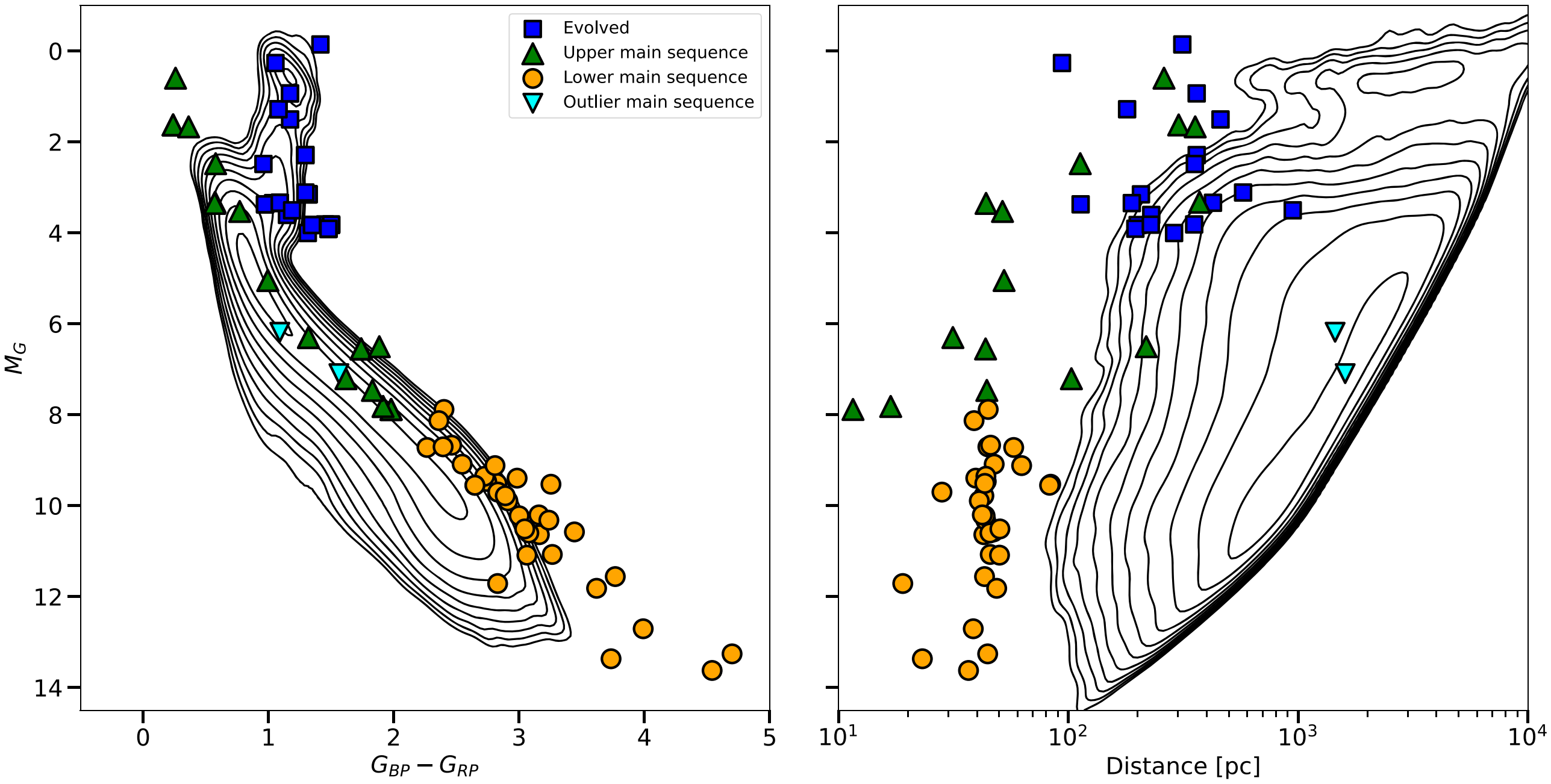

**Figure 5.** Left: CMD of SPT flaring stars, with contours showing density of Gaia stars in the SPT-3G winter field. Right: magnitude vs. distance of the same populations. The two inverted triangles are outlier main-sequence stars and considered to be incorrect matches, which are discussed in Section 3.8.3.

M dwarf stars, having a Gaia $G_{BP} - G_{RP}$ color of $\gtrsim$2.1, are interesting, due to both their tendency for large and frequent flares (France et al. 2016) and their abundance—accounting for ~75% of stars in our local neighborhood (Bochanski et al. 2010). The M dwarf population that the SPT has seen flare is largely above the main sequence, indicating that these could be young stars that have not yet settled onto the main sequence.

Early-type stars (spectral types B5−F5 with a Gaia $G_{BP} - G_{RP}$ color of $\lesssim$0.5) are also represented in the SPT flaring population. While initially believed to not exhibit flares, there has been recent evidence of flaring activity in early-type stars at other wavelengths (Balona 2012; Van Doorsselaere et al. 2017). The upper main-sequence stars that have been detected flaring by SPT are generally closer in distance than typical main-sequence stars. This could indicate that flaring in these types of stars could be very common, and SPT is preferentially detecting the sample that happens to be nearby.

### 3.3. Spectral Index

Spectral indices for the events are shown in the main plot in Figure 6. Events are separated by their 220 GHz signal-to-noise ratio (SNR), with a focus on events with strong (>3$\sigma$) detections. Events with weak 220 GHz SNR have their $y$-error bars omitted and marker sizes reduced for visual clarity.

The spectral indices are defined through the relation

$$S_\nu \propto \nu^\alpha, \tag{6}$$

such that

$$\alpha_{\nu_2}^{\nu_1} = \frac{\log(S_1/S_2)}{\log(\nu_1/\nu_2)}. \tag{7}$$

To represent probability densities of the observed spectral indices, we have calculated kernel density estimations (KDEs) for two populations: a KDE for the subset of events with strong 220 GHz detections, and a KDE for the entire population of flaring events regardless of 220 GHz SNR, in both $\alpha_{150}^{95}$ and $\alpha_{220}^{150}$. Gaussian kernels are used with the width set by the uncertainty on the spectral index measurement. We show these KDEs along the $x$- and $y$-insets, as well as median spectral index values for the two populations.

Most of the events detected by SPT cluster around the constant spectral index line, with median spectral indices of $\alpha \approx 0$–0.5. Measurements of $\alpha$ generally do not exceed 5/2, which is the spectral index for optically thick synchrotron radiation (Pacholczyk 1970), but most do not seem consistent with a falling spectrum expected for the optically thin side of synchrotron radiation that peaked at radio wavelengths around ~5 GHz.

In the flux subplots, spectral indices with typical values in the radio regime are marked with black lines. The value $\alpha = -0.75$ represents an optically thin synchrotron source, $\alpha = 0$ represents a flat spectral index, $\alpha = 2$ represents the Rayleigh–Jeans tail of thermal emission, and $\alpha = 5/2$ represents an optically thick synchrotron source due to self-absorption. The SPT-detected events fall within the bounds of typical synchrotron spectra for both $\alpha_{150}^{95}$ and $\alpha_{220}^{150}$, suggesting this as a plausible emission mechanism.

### 3.4. Number Counts

Figure 7 shows cumulative number counts of both flux density and 150 GHz peak flare luminosity, split for the three populations of stars we have established.

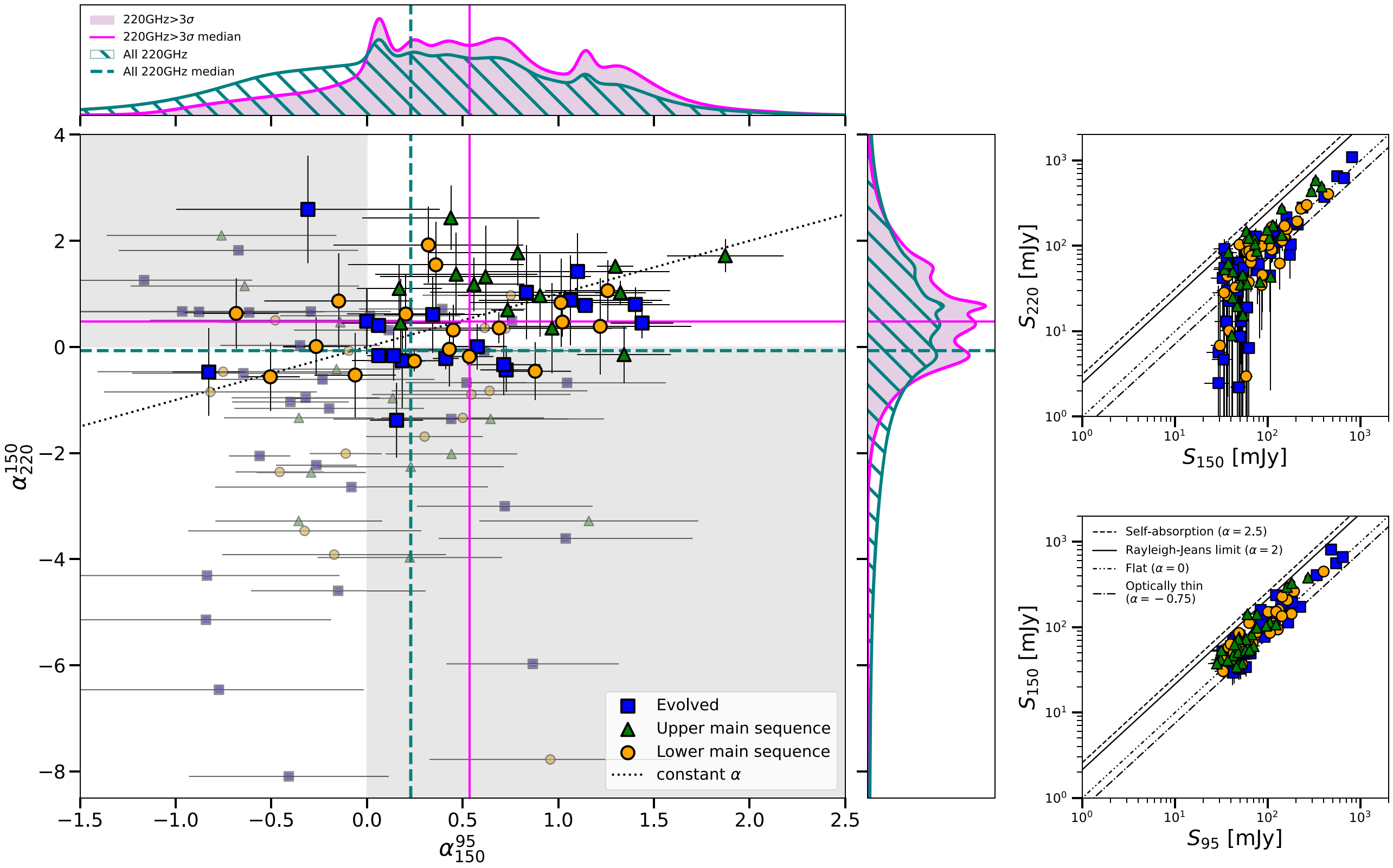

**Figure 6.** Left: spectral indices for all stellar flare events, with KDEs of the probability densities along the insets. Shaded regions mark separate quadrants of positive and negative values of alpha, while the dotted line indicates a spectral index that is constant in frequency. Events with 220 GHz SNR $< 3\sigma$ have their $y$-error bars omitted and marker sizes reduced for visual clarity. These spectral indices are within the typical values for synchrotron radiation more commonly seen at radio wavelengths. Right: comparisons of flux densities for all stellar flare events, across the three SPT bands. Typical spectral indices of relevant phenomena are shown as lines.

The flux density number count can be approximated as a power law of $N(>S) \approx S^{-3/2}$, where the index comes from assuming a uniform density of objects within a sphere: $N \propto r^3$ and $S \propto r^{-2}$. This approximation is shown as a series of dashed lines in the flux density number count plot. At distances that exceed the thickness of the galactic disk, the power-law approximation will break down. The lower main-sequence stars are <100 pc in distance and are not detected outside of the disk. However, the evolved stars are seen at distances up to 1 kpc and may be detected at distances where a drop in density is expected.

The luminosity number count shows that the main-sequence stars have similar luminosity distributions, while the evolved stars have much brighter flares and higher flaring rates in comparison. Incorporating the transient detection efficiency results of flares lasting 30 minutes or less as discussed in Section 2.1, we find an expected flaring rate $N(>L)$ deg$^{-2}$ yr$^{-1}$ to be within a factor of 2 of 0.2/0.3/0.4 for the upper main-sequence/lower main-sequence/evolved populations of stars at a $\nu L_\nu$ 150 GHz luminosity of $10^{27}$ erg s$^{-1}$.

### 3.5. X-Ray

For spectral type F to M stars, the Güdel–Benz relation (Guedel & Benz 1993) gives an empirical linear relation between radio emission at 5 GHz and thermal SXR emission at a wide range of energies, directly relating emission from the particle acceleration phase to the coronal heating. This results in the observed correlation $\log L_R \approx \log L_X - 15.5$, where $L_R$ is the radio luminosity and has units of erg s$^{-1}$ Hz$^{-1}$ and $L_X$ is the SXR luminosity and has units of erg s$^{-1}$.

We are unable to include a suitable comparison for millimeter-wave emission detected by SPT, due to the mixing of flaring and quiescent states. Only two stars are detected above $3\sigma$ levels in our 3 yr average flux: $2.47 \pm 0.70$ mJy and $3.31 \pm 0.68$ mJy at 150 GHz. This averaged flux is a combination of detected flares and any weaker flares below our detection threshold and is considered to be an unreliable upper limit on flux. We are also unable to supplement our data with radio emission from other surveys, with the Rapid ASKAP Continuum Survey (McConnell et al. 2020) only having a small number of sources within close proximity (30″) of SPT flaring stars.

An investigation of quiescent SXR emission is still useful, however. Studies of the Sun and other active main-sequence stars have shown that a linear relationship exists between SXR emission and magnetic flux (Pevtsov et al. 2003), linking thermal coronal emission to magnetic fields. Flaring rates and luminosity have been linked to quiescent SXR emission (Audard et al. 2000), as well as other proxies for stellar magnetic activity such as Hα emission and rotation period (Davenport 2016; Yang et al. 2017).

Figure 8 shows the SPT flaring stars at 150 GHz, along with quiescent SXR (0.1–2.4 keV) from 2RXS in luminosity, using associations provided by Freund et al. (2022); the right panel shows a histogram of the ratios of luminosities. Four sources not matched in Freund et al. (2022) but within 1′ of sources included in Boller et al. (2016) are shown with a dashed circle around the marker. Their flux ($F_X$) is calculated following Schmitt et al. (1995) using the hardness ratio (HR) and photon count rate (ctr):

$$F_X = (5.30\ \mathrm{HR} + 8.31)\ \mathrm{ctr} \times 10^{-12}\ \mathrm{erg\ cm^{-2}\ s^{-1}}. \quad (8)$$

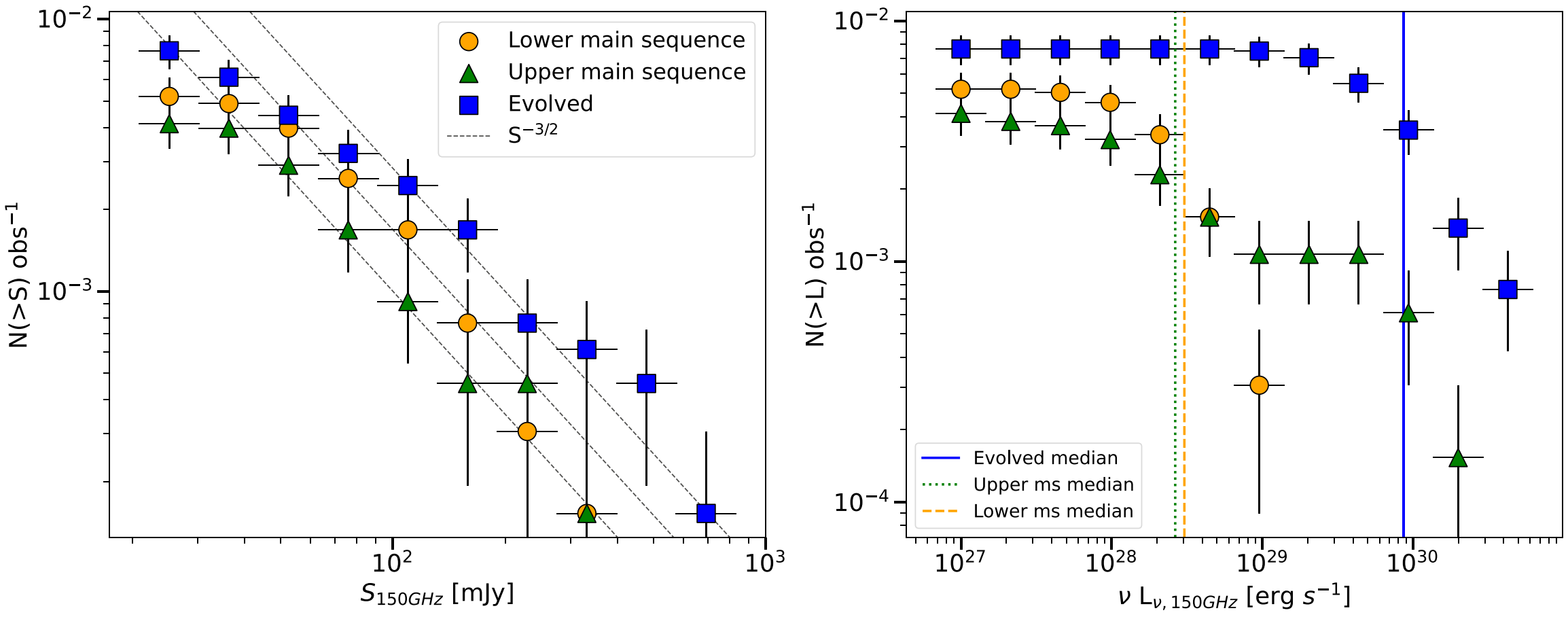

**Figure 7.** Cumulative number counts per SPT observation for flaring events at 150 GHz. Left: number counts as a function of source flux densities. Dashed lines are examples of the Euclidean source count expected power law of $S^{-3/2}$. Right: number counts as a function of peak flare luminosities. Vertical lines show the median luminosity value for each population of stars.

We have used this equation to recover the same flux values for sources given in Freund et al. (2022), using the HRs and photon count rates supplied within.

Considering only the SPT-detected flare stars with an association in 2RXS (i.e., excluding the matches within $1'$ proximity), the percentage of stars that are X-ray bright is 54.5%. For comparison, we select a population of 330 stars that are representative of the SPT flaring star population. Using the parameters of magnitude and distance (see the right panel of Figure 5), we selected five nearest neighbors for each SPT flare star. The nearest neighbor stars have a percentage of X-ray-bright stars of 4.8%. These rates provide evidence that stars observed to flare at millimeter wavelengths are much more likely to be X-ray bright than nonflaring stars.

The histograms in Figure 8 show a large scatter in the ratio of quiescent SXR luminosity to millimeter-wave flaring luminosity for the population of stars with 2RXS associations. Simultaneous observations of SXR and millimeter-wave emission during flaring states may reveal a tighter correlation between these luminosities.

### 3.6. Single-scan Light Curves

Flare morphologies on short timescales can tell us about the physics of the flare. However, binning or gaps in data can obscure important features of these flares. At optical wavelengths, higher-cadence observations from TESS have shown a degeneracy in data even at the 2-minute scale, with subsequent pulsations after the peak and during the decay phase possible in a larger portion of the population (Howard & MacGregor 2022).

Light curves at radio wavelengths typically describe the impulsive phase of the flare, but a phase of gradual change at these wavelengths has also been detected (Zhang et al. 2018). The impulsive stellar flare light curve seen in radio, optical, and HXR wavelengths is characterized by a sharp rise followed by an exponential decay, with some small complexities occurring during both phases (Davenport et al. 2014; Kowalski et al. 2019). Different types of stars can also exhibit unusual flare morphologies: flares from RS CVn binary systems have longer rise times and sustained peak emission that plateaus before decaying (Osten & Brown 1999).

Flaring activity can also trigger additional flares in neighboring active regions of the star in a process known as sympathetic flaring. Sympathetic flaring has been observed on the Sun (Pearce & Harrison 1990), with evidence of this phenomena happening in other stars (Anfinogentov et al. 2013). This could explain the appearance of a complex flare as the superposition of separate flares on the surface of the star.

Most descriptions of the duration of stellar flares at shorter wavelengths benefit from higher-cadence observations, longer times spent viewing the star, or some combination of both. Due to the SPT scanning strategy, we are limited to the binned single-scan data points, roughly 100 s apart on average, with the actual time spent on a given sky location depending on where that location falls in the focal plane's approximately oval shape. As these are not targeted observations, we rarely observe the duration of a flare in its entirety. Such observations can only give us a lower limit on the duration of these events.

Figure 9 shows light curves of two events with clear morphologies, including fits of a long Gaussian rise time and an exponential decay using `scipy.optimize.curve_fit`. Visual inspection resulted in around 15 detected flares being considered to have clear structure, while most flares did not have a well-defined evolution of the light curve during the time observed. Additional single-scan light curves for all SPT stars are included in the Appendix; these light curves also contain 220 GHz data, which have approximately 5× higher noise than those in the 95 and 150 GHz bands.

The peak flux values of flares reported in the catalog are averaged over full observations and thus have a lower peak flux value than the corresponding single-scan light curve. Shorter-duration events have larger biases since typical observations contain 16 single-scan data points. An event that is shorter than 100 s in duration may only be caught in one single scan and will be reduced by a factor of 16 when using the single-observation flux instead. We choose to represent the stellar flares with single-observation flux values, due to our transient detection pipeline running on single-observation light curves, as well as the increased noise values inherent in the single-scan data. Figure 10 shows the difference in 150 GHz luminosities for flares using peak single-observation and peak single-scan fluxes, showing that the average difference in recovered luminosity is roughly a factor of 2.

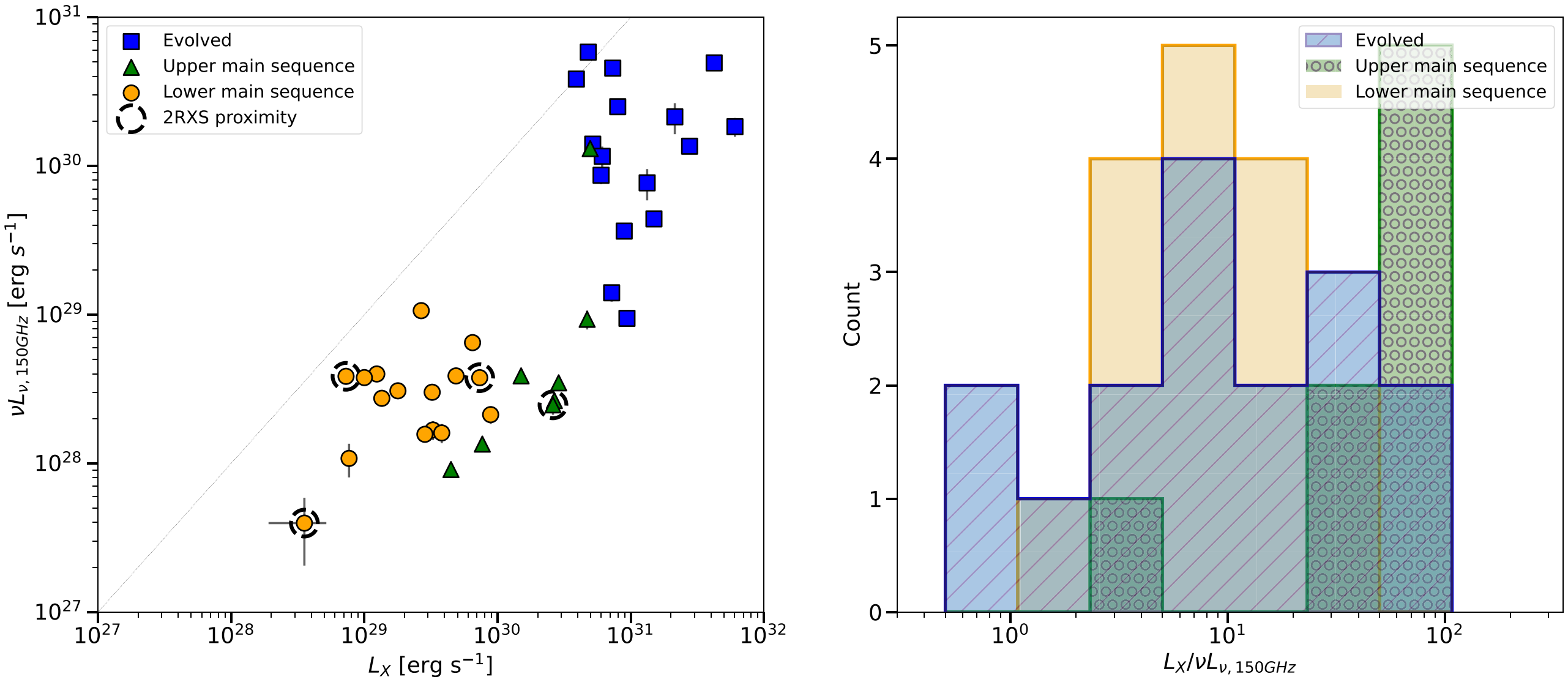

**Figure 8.** Left: comparison of millimeter flaring and X-ray quiescent luminosities. A dotted line shows equal luminosities. Right: a histogram of luminosity ratios for the three populations of stars that have 2RXS associations through Freund et al. (2022) and proximity to objects in Boller et al. (2016). These plots show a wide spread in the ratio of quiescent SXR to flaring millimeter-wave luminosities.

Our single-scan light curves occasionally have additional peaks in them. When considering the typical duration for the impulsive phase of a stellar flare on the order of seconds, we cannot confidently classify these peaks as a separate event occurring nearby temporally, or a pre/post-flare pulsation as described in Howard & MacGregor (2022), where similarly binned TESS data at 120 s cadence are shown to remove features in M dwarf flare morphologies that are apparent when viewing the 20 s cadence light curve of the same flare.

### 3.7. Flaring Activity Durations

We constrain the duration of flaring activity by using the single-scan and single-observation light curves for each event. Lower limits on duration are set by the single-scan light curve, with a flare considered to be at quiescence (either before or after the peak) by having a single-scan SNR below $4\sigma$ simultaneously in both bands. For a flare to be considered fully contained within a single-scan light curve, quiescent points need to exist on either side of the flare. For flares with only one quiescent point in their single-scan light curve, the minimum duration is considered to be from this quiescent point to either the beginning or end of the light curve depending on where it is in relation to the peak. Flares with no quiescent points in their single-scan light curve use the full time between the first and last scan as a minimum duration. Upper limits on duration are constrained in a similar way: quiescence is considered to be when the closest single observation returns to $1\sigma$ before and after the peak.

Figure 11 shows the constraints on duration for all flaring events. We find five flares to be fully constrained within the single-scan light curve and have measured their durations: these are plotted as larger marker sizes with an uncertainty of one full single scan (100 s). The shaded area denotes the range in possible times that a location will be observed by the SPT, from one scan (100 s) to 20 scans, including the time for the SPT to accelerate and change elevation while it is not collecting data (2060 s). The vertical line shows the typical duration of a full single observation (7668 s).

Going from single-observation to single-scan light curves increases our time resolution but still suffers a low cadence, as there are 100 s on average between data points. Depending on where in the subfield the flaring event occurs, subsequent single-scan data points may represent just a few seconds of the focal plane passing over that part of the sky (see Figure 3), with large gaps in time possible, as it takes the telescope 100 s to complete a scan in one direction. This cadence is more than sufficient to hide features in a flare.

Similarly, large gaps between single-observation data points can span anywhere from 2 to 24+ hr depending on the subfield observation schedule. For most flares, these temporal gaps in SPT data are large enough that it is likely that the true duration of the flare is much closer to the lower limit. Some flares exhibit multiple single-observation data points that are likely part of the same long-duration flare but could also possibly be separate back-to-back flares. The transient detection pipeline only fits for a single flare in each 12-day light curve, so additional flares in the same light curve will go unreported. Light curves for all events have been manually analyzed for obvious secondary flares, and none have been found. This fitting method will be improved on in the future to allow for finding multiple flares within the same light curve.

### 3.8. Additional Classifications

#### 3.8.1. Binary Systems

We have checked for binarity in the sample of SPT-detected flare stars. Binary systems can lead to confusion in determining the correct source—there is a possibility of matching to the wrong star, due to undetected secondaries or preferentially selecting the brightest star in Gaia. Stars in binary systems can also exhibit flares due to interactions between the stars, such as RS CVn systems that have been detected at millimeter

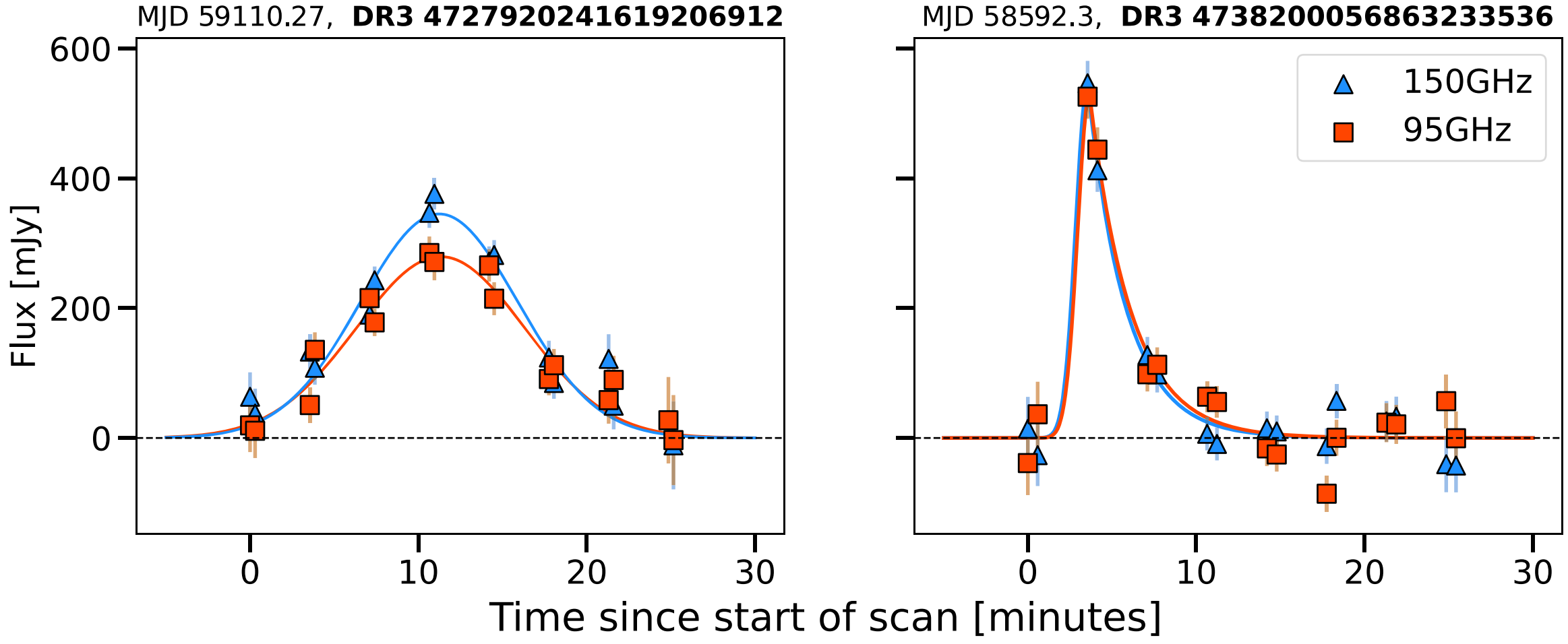

**Figure 9.** Possible stellar flare models on well-constrained single-scan light curves. As shown in the Appendix, most flares detected by SPT are not as clear in their structure. Both of these light curves are of flares from lower main-sequence stars. Left: a simple Gaussian model showing a gradual rise in flux that is comparable to the decay period. Right: a sharp Gaussian rise with a longer exponential decay.

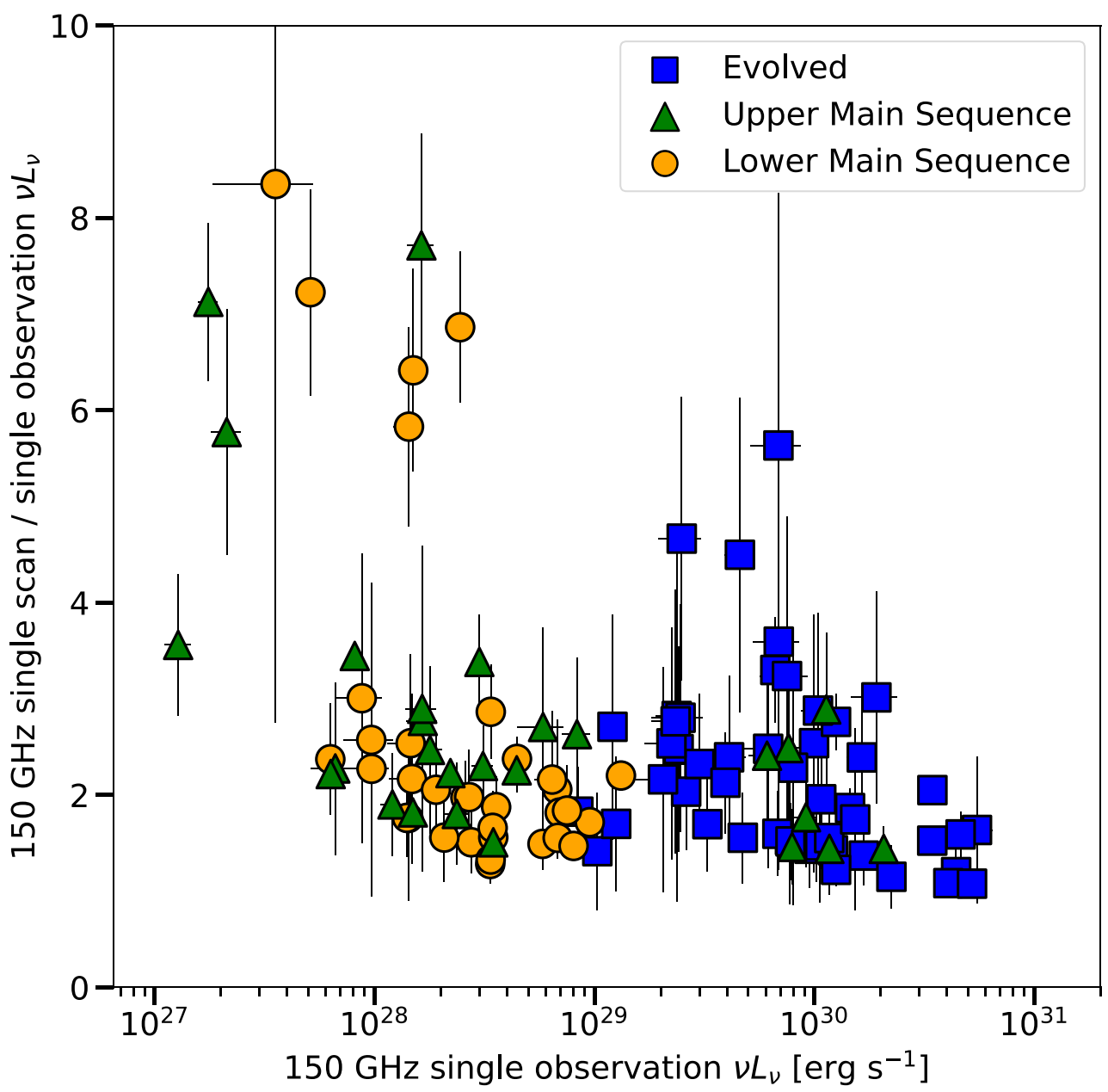

**Figure 10.** Comparison of single-scan and single-observation peak 150 GHz luminosities for each event. The flux values from observations are a factor of 2 lower than the peak from single scans for the majority of flares.

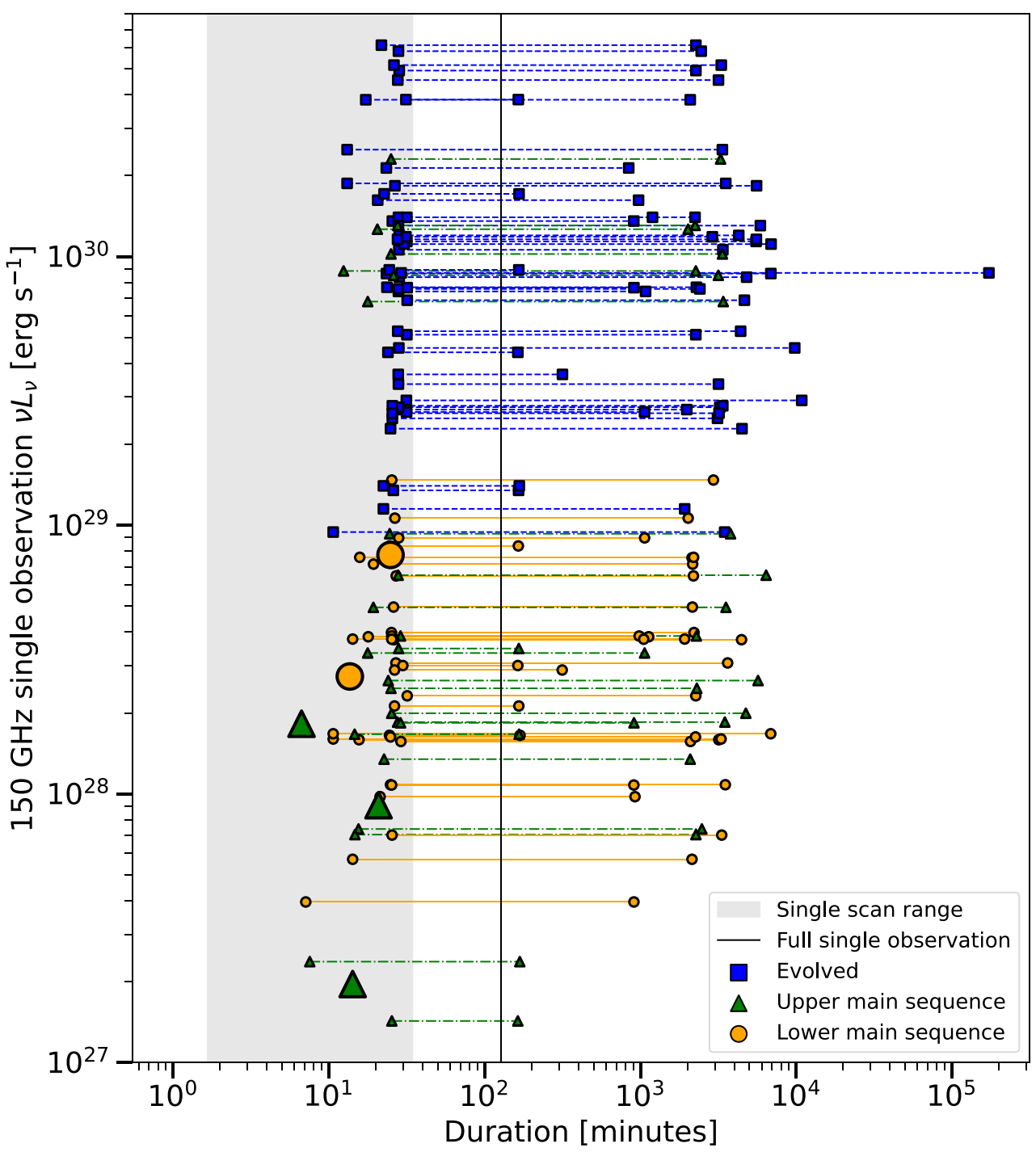

**Figure 11.** Constraints on durations for stellar flares, using single-scan light curves to determine lower limits and single-observation light curves to determine upper limits. Five events are found to be fully contained within the single-scan light curve and have single markers at a larger size.

wavelengths (Brown & Brown 2006) and are known to flare frequently (Osten & Brown 1999).

Using Vizier (Ochsenbein et al. 2000), we have searched all flaring stars for inclusion in catalogs of binary stars, as well as classification by Gaia (Rimoldini et al. 2023). We use the same process described in Section 3.5 to compare SPT-detected stars with a similar representative population in Gaia.

We find that the SPT population of stars is overrepresented in RS CVn classification by Gaia (10.6% compared to 0.6% for the nearest neighbor population), as well as probable binary candidates (51.5% compared to 22.7%). While many of these classifications require more definitive proof, such as spectroscopic observations, the higher rates of binarity are in agreement with studies showing increased flaring rates in binary systems (Huang et al. 2020; Yang et al. 2023).

### 3.8.2. Multiple Flares

Figure 12 shows a CMD of the stars we have detected flaring multiple times. While these stars sample the entirety of our population, we find that our detections of multiple flares are dominated by evolved stars, with 30 of the 62 events coming from just five stars. From preliminary simulations of injected flare efficiency (Section 2.1), it is likely that the true rate of above-threshold flares from these stars ranges from 2× to 20× higher than observed. Figure 11 shows only five flares—from upper and lower main-sequence stars—to be fully constrained and to have durations on the scale of 30 minutes or less, suggesting that the main-sequence stars in Figure 12 have a true flaring rate closer to 20× higher. Table A1 shows a full listing of these stars and all of their associated events.

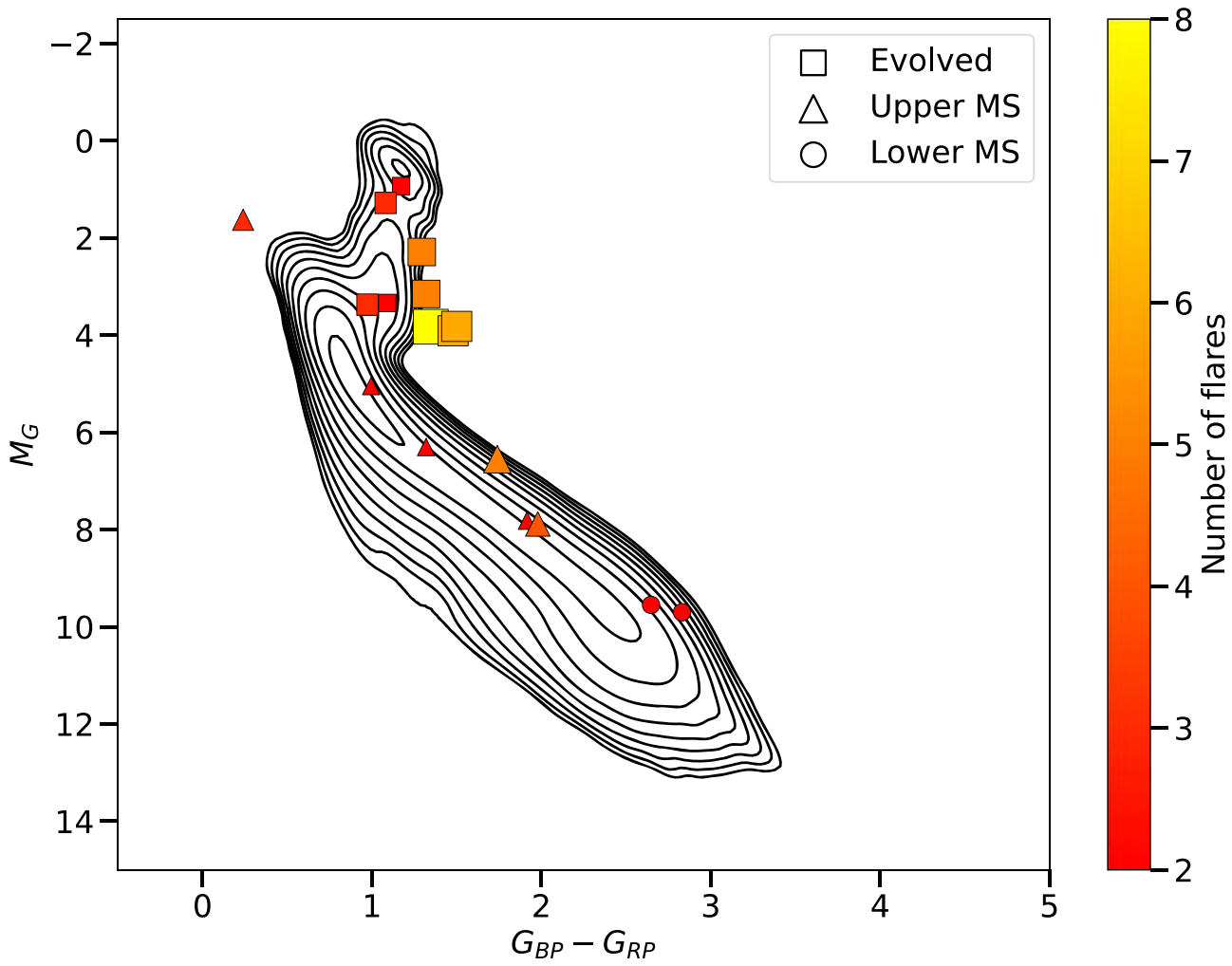

**Figure 12.** Gaia CMD showing stars that have multiple flares detected by the SPT with number of flares shown both by color and scaling of marker size. Most of the multiple flares detected originate from evolved stars.

### 3.8.3. Outlier Events

Flaring events in Figure 4 having $p$-values greater than 0.02 have been individually investigated. In most cases the higher-than-typical $p$-value is due to the associated star being in a high local density region and having multiple nearby candidate stars, and we believe these matches to be correct.

For the two events on MJD 58706.43 ($p$: 0.038) and MJD 59767.07 ($p$: 0.086), our cross-matching algorithm picked a dimmer star closer in angular separation rather than brighter stars at slightly larger separation. In addition, both events are matched to extreme outliers: they are main-sequence stars at distances over 1 kpc, significantly farther away than other stars we match to. Figure 5 shows these two anomalous stars in the middle of the main sequence as having much farther distances than similar main-sequence stars the SPT has detected, resulting in extremely high flaring luminosities. For both of these events, the other potential candidates are nearby M dwarfs: the candidate for MJD 58706.43 is a potential binary system (Reiners et al. 2010), and the two candidates for MJD 59767.07 have an angular separation of 6.″2 and distances of 51.13 and 51.17 pc, giving a distance from each other of ∼0.04 pc. We believe that the M dwarfs are a more likely match for both events, and we have changed the associated sources.

These cases suggest that some improvements in our cross-matching algorithm can be made. The events were preferentially matched to a distant main-sequence star, rather than a brighter M dwarf at slightly higher angular separation; it is possible that including a prior on parallax based on our confirmed distributions in the CMD can steer the event to a more plausible match. Likewise, incorporating a prior that favors stars that have been observed flaring previously might be considered. Further optimization of this algorithm will be necessary to inform future analysis of surveys such as CMB-S4 and SO, and especially for surveys that include the galactic plane.

## 4. Conclusion

We have presented a new catalog of short-duration (<10 days) transient events found in a nontargeted SPT-3G survey using an updated transient detection and matching method. This catalog is produced from analyzing 4 yr of daily observations of a 1500 deg$^2$ field at 95 and 150 GHz by the SPT-3G survey, with the inclusion of 220 GHz data coming after events have been detected. The catalog is provided in machine-readable format at https://pole.uchicago.edu/public/data/tandoi24/, along with 4 yr single-observation light curves at 95 and 150 GHz of each unique star and single-scan light curves at 95, 150, and 220 GHz of each flaring event at 10.5281/zenodo.11372033.

We have found that two-band spectral index comparisons for 95/150 GHz and 150/220 GHz generally fall between an optically thick synchrotron spectrum ($\alpha = 5/2$) and a typical value of the optically thin tail of synchrotron ($\alpha = -0.75$). While optically thick synchrotron radiation typically peaks at radio frequencies (∼5 GHz), evidence of rising spectra at submillimeter frequencies has been observed in solar flares.

We have found these flaring events to be nearby stars that we have matched to Gaia Data Release 3. Many of these stars have counterparts in the 2RXS stellar content catalog. A CMD shows our population of matched stars, indicating that we are seeing flaring events from stars spanning a wide range of spectral types, with most of these flares occurring in M dwarfs and along the giant branch.

Quiescent SXR emission has been shown to have a linear relationship with magnetic flux in main-sequence stars, assumed to be the cause for the flaring activity we have observed, along with measures of direct flaring activity. We have shown quiescent SXR emission along with peak flare 150 GHz luminosity, as well as a comparison of this ratio, and found strong evidence that millimeter-wave flaring stars tend to be quiescently bright in SXR. A wide spread in this ratio exists that may be tightened up with simultaneous SXR and millimeter emissions during a flare.

We have shown cumulative number counts of the flaring stars in flux densities and peak flare luminosities. The flux density number count has a slope consistent with observing sources in a uniform density, i.e., the Milky Way disk, at higher flux values, with this assumed power law of $N(>S) \approx S^{-3/2}$ breaking as flux density approaches the SPT transient detection threshold. The number count as a function of luminosity shows lower and upper main-sequence stars having similar median luminosities. Evolved stars have a median peak flare luminosity over 1 mag brighter than the SPT-detected main-sequence stars.

We have flagged all probable binary system candidates, known through Gaia variability tables and a literature search in Vizier. It is not clear whether these flares originate from the associated star, a companion, or an interaction between the stars in the system.

We have observed 17 out of 66 stars flare multiple times, comprising over half of our events and a wide range of spectral types, with most of these multiflare events originating from just a few evolved stars.

We believe that our cross-matching algorithm has erroneously matched two events with distant stars, resulting in anomalously high luminosities. Both matches are to main-sequence stars at distances of >1 kpc (greater than any other Gaia star that SPT has observed flaring) and are closer in angular separation to the SPT event compared to nearby M dwarfs at ≈50 pc, which are brighter in apparent magnitude. We believe that the M dwarfs are the source of the flaring activity for both of these events. In the future, we are considering adding a prior based on location in the Gaia CMD to our cross-matching algorithm.

We have shown finer time resolution single-scan light curves of events, using these as a lower-limit constraint on durations. Upper limits on the duration are found using single-observation light curves, and we have found typical constraints on duration ranging from $\sim 10^1$ to $\sim 10^4$ minutes. These constraints are weak and are generally dominated by the SPT observation cadence. Two potential models are shown to fit to some of the shorter-duration single-scan light curves: a sharp rise with exponential decay, and a Gaussian fit with a rise time comparable to the decay time.

The stellar flares we have detected with the SPT are brighter and longer in duration than flares detected by ALMA at similar wavelengths, which in turn are brighter than the brightest solar flares detected. It is possible that these are all parts of the same flaring distribution, but a lack of polarization, along with a wide range of spectral types and the possibility of binary systems, means that we are unable to confidently assign a single emission mechanism as the cause of flaring among these populations.

Simultaneous observations will be possible for future SPT observing seasons, with TESS having revisited the SPT-3G field in 2023 July, and with the ASKAP Variables and Slow Transients (VAST) survey covering the same area of the sky over parts of the next few years. With additional wavelengths, as well as polarization in the case of VAST, these surveys will deliver important information as to the nature of these flaring events. Targeted observations of known multiflaring SPT stars from radio instruments such as ALMA in conjunction with the SPT can extend the radio SED and provide the finer time resolution that can constrain the durations of these events. We can also use SPT-3G data to search for linearly polarized emission.

The flaring events in this catalog have greatly increased the known number of millimeter-wavelength flare stars. Additionally, the methods used to create this catalog—map filtering, transient detection, flare efficiency simulations, and cross-matching to Gaia—have built uon previous SPT transient analysis and will continue to be improved. These methods and the resulting catalog from 4 yr of the nontargeted SPT-3G survey will form a basis for what to expect while searching for short-duration transient events and are important for planning future millimeter-wavelength surveys such as CMB-S4 and SO.

## Acknowledgments

The authors thank Rachel Osten, Alex Gagliano, Konstantin Malanchev, Gautham Narayan, and Brian Fields for helpful discussions. The South Pole Telescope program is supported by the National Science Foundation (NSF) through award OPP-1852617. Partial support is also provided by the Kavli Institute of Cosmological Physics at the University of Chicago. Work at Argonne National Lab is supported by UChicago Argonne LLC, Operator of Argonne National Laboratory (Argonne). Argonne, a US Department of Energy Office of Science Laboratory, is operated under contract No. DE-AC02-06CH11357. Work at Fermi National Accelerator Laboratory, a DOE-OS, HEP User Facility managed by the Fermi Research Alliance, LLC, was supported under contract No. DE-AC02-07CH11359. The IAP authors acknowledge support from the Centre National d'Études Spatiales (CNES). This project has received funding from the European Research Council (ERC) under the European Union's Horizon 2020 research and innovation program (grant agreement No. 101001897). The Melbourne authors acknowledge support from the Australian Research Council's Discovery Project scheme (No. DP210102386). The McGill authors acknowledge funding from the Natural Sciences and Engineering Research Council of Canada, Canadian Institute for Advanced Research, and the Fonds de recherche du Québec Nature et technologies. The SLAC authors acknowledge support by the Department of Energy, contract DE-AC02-76SF00515. The UCLA and MSU authors acknowledge support from NSF AST-1716965 and CSSI-1835865. M.A., K.A.P., J.V., and Y.W. acknowledge support from the Center for AstroPhysical Surveys at the National Center for Supercomputing Applications in Urbana, Illinois. K.F. acknowledges support from the Department of Energy Office of Science Graduate Student Research (SCGSR) Program. Z.P. was supported by Laboratory Directed Research and Development (LDRD) funding from Argonne National Laboratory, provided by the Director, Office of Science, of the US Department of Energy under contract No. LDRD-2021-0186. J.V. acknowledges support from the Sloan Foundation. W.L.K.W is supported in part by the Department of Energy, Laboratory Directed Research and Development program and as part of the Panofsky Fellowship program at SLAC National Accelerator Laboratory, under contract DE-AC02-76SF00515. This research has made use of the SIMBAD database, operated at CDS, Strasbourg, France. This research has made use of the VizieR catalog access tool, CDS, Strasbourg, France (DOI:10.26093/cds/vizier). The original description of the VizieR service was published in Ochsenbein et al. (2000). This research was done using services provided by the OSG Consortium (Pordes et al. 2007; Sfiligoi et al. 2009; OSG 2006, 2015), which is supported by the National Science Foundation awards #2030508 and #1836650. The data analysis pipeline also uses the scientific Python stack (Jones et al. 2001; Hunter 2007; van der Walt et al. 2011). This work has made use of data from the European Space Agency (ESA) mission Gaia (https://www.cosmos.esa.int/gaia), processed by the Gaia Data Processing and Analysis Consortium (DPAC, https://www.cosmos.esa.int/web/gaia/dpac/consortium). Funding for the DPAC has been provided by national institutions, in particular the institutions participating in the Gaia Multilateral Agreement. This publication makes use of data products from the Wide-field Infrared Survey Explorer, which is a joint project of the University of California, Los Angeles and the Jet Propulsion Laboratory/California Institute of Technology, and NEOWISE, which is a project of the Jet Propulsion Laboratory/California Institute of Technology. WISE and NEOWISE are funded by the National Aeronautics and Space Administration.

*Facilities:* SPT, Gaia, ROSAT, WISE

## Appendix

Figure A1 shows single-scan light curves for each SPT flaring event. Limits on the *y*-axis are set to highlight the morphology of the 95 and 150 GHz data. The consequence of this is that some 220 GHz data may approach the edge of the figure and appear to be cut off. Table A1 shows luminosities and spectral indices for all stars that were detected flaring multiple times.

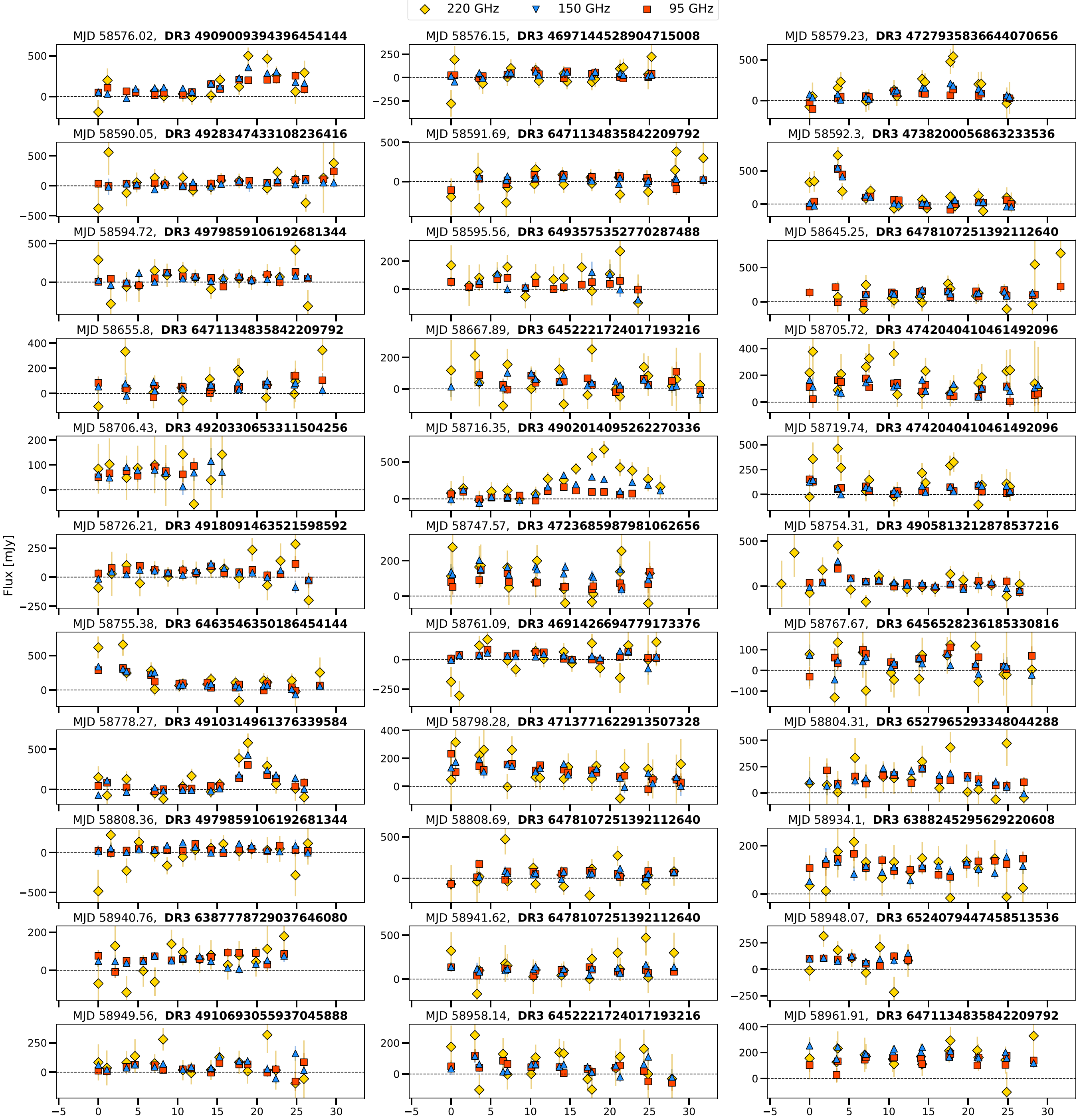

**Figure A1.** Additional single-scan light curves.

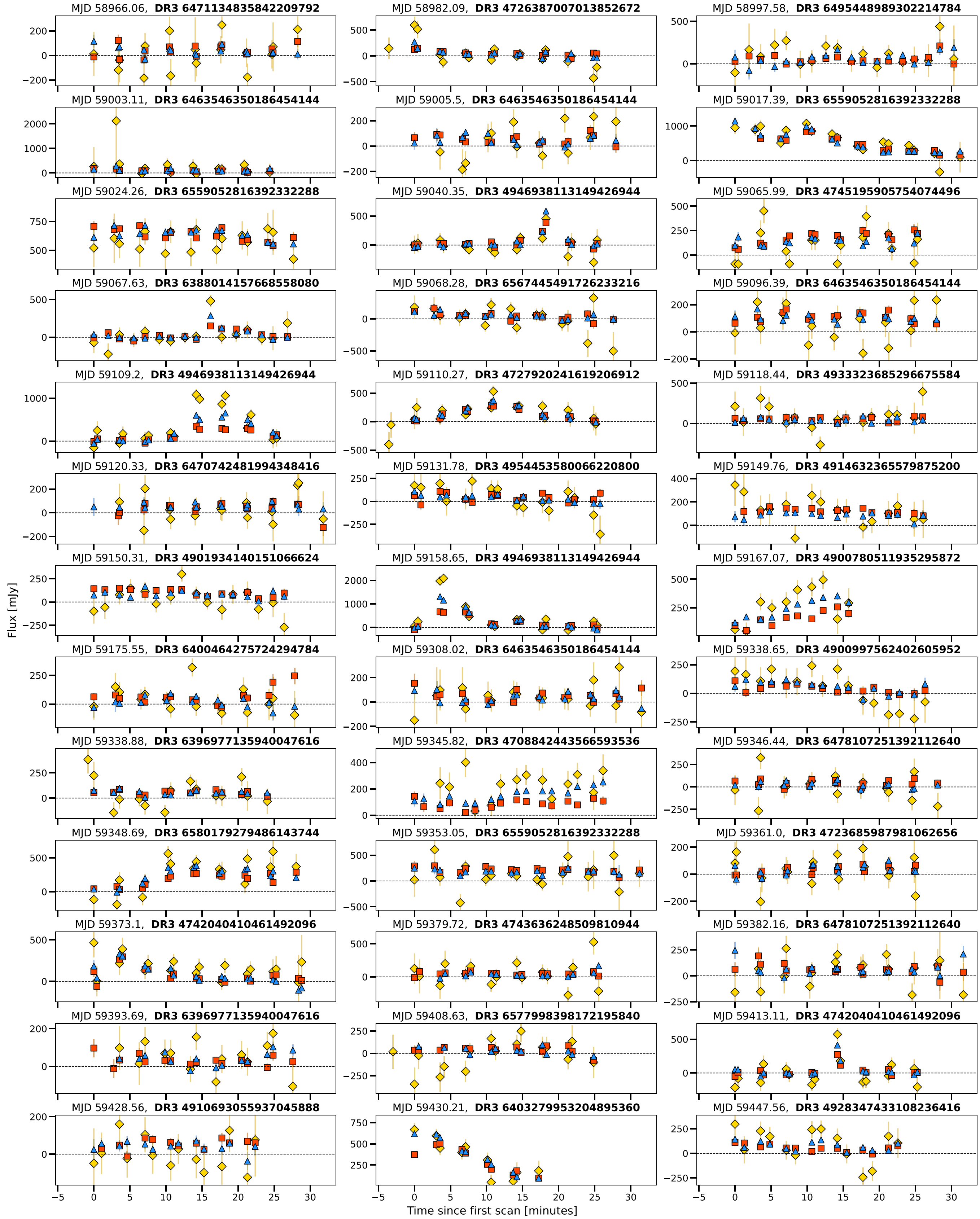

**Figure A1.** (Continued.)

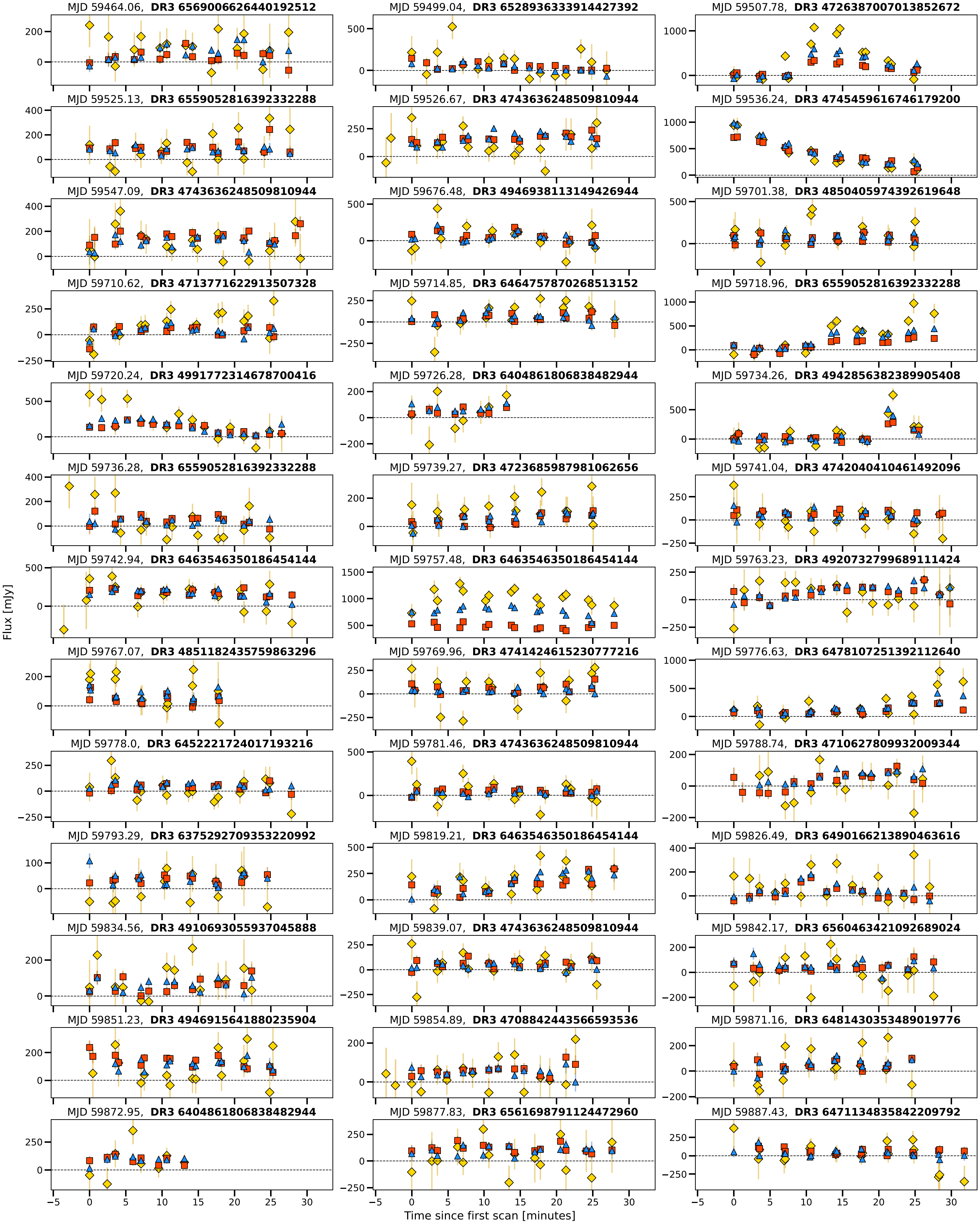

**Figure A1.** (Continued.)

**Table A1**
Luminosities and Spectral Indices for All Stars with Multiple Detected Flares

| Gaia DR3 ID | MJD | $\nu L_{\nu,95}$ (erg s$^{-1}$) | $\nu L_{\nu,150}$ (erg s$^{-1}$) | $\nu L_{\nu,220}$ (erg s$^{-1}$) | $\alpha^{95}_{150}$ | $\alpha^{150}_{220}$ | Flares |
|---|---|---|---|---|---|---|---|
| 6463546350186454144 | 59757.48 | $2.19 \times 10^{30}$ | $5.81 \times 10^{30}$ | $1.15 \times 10^{31}$ | $1.14 \pm 0.04$ | $0.78 \pm 0.07$ | 8 |
| | 59742.94 | $8.32 \times 10^{29}$ | $1.40 \times 10^{30}$ | $1.93 \times 10^{30}$ | $0.14 \pm 0.14$ | $-0.17 \pm 0.47$ | |
| | 59819.21 | $6.34 \times 10^{29}$ | $1.30 \times 10^{30}$ | $1.92 \times 10^{30}$ | $0.58 \pm 0.15$ | $0.0 \pm 0.43$ | |
| | 59003.11 | $4.00 \times 10^{29}$ | $8.92 \times 10^{29}$ | $1.58 \times 10^{30}$ | $0.76 \pm 0.35$ | $0.49 \pm 1.01$ | |
| | 59096.39 | $5.42 \times 10^{29}$ | $7.58 \times 10^{29}$ | $4.73 \times 10^{29}$ | $-0.27 \pm 0.21$ | $-2.23 \pm 1.63$ | |
| | 58755.38 | $4.70 \times 10^{29}$ | $7.42 \times 10^{29}$ | $1.31 \times 10^{30}$ | $0.0 \pm 0.24$ | $0.48 \pm 0.61$ | |
| | 59005.50 | $2.32 \times 10^{29}$ | $3.35 \times 10^{29}$ | $3.15 \times 10^{29}$ | $-0.2 \pm 0.5$ | $-1.16 \pm 2.39$ | |
| | 59308.02 | $1.86 \times 10^{29}$ | $2.64 \times 10^{29}$ | $3.06 \times 10^{29}$ | $-0.23 \pm 0.59$ | $-0.61 \pm 2.51$ | |
| 6559052816392332288 | 59024.26 | $2.79 \times 10^{30}$ | $4.54 \times 10^{30}$ | $6.25 \times 10^{30}$ | $0.06 \pm 0.04$ | $-0.16 \pm 0.14$ | 6 |
| | 59017.39 | $2.37 \times 10^{30}$ | $3.84 \times 10^{30}$ | $6.58 \times 10^{30}$ | $0.06 \pm 0.04$ | $0.41 \pm 0.13$ | |
| | 59718.96 | $5.32 \times 10^{29}$ | $1.62 \times 10^{30}$ | $2.83 \times 10^{30}$ | $1.44 \pm 0.16$ | $0.45 \pm 0.28$ | |
| | 59353.05 | $9.69 \times 10^{29}$ | $1.18 \times 10^{30}$ | $7.91 \times 10^{29}$ | $-0.56 \pm 0.16$ | $-2.06 \pm 1.25$ | |
| | 59525.13 | $4.01 \times 10^{29}$ | $5.27 \times 10^{29}$ | $5.20 \times 10^{29}$ | $-0.4 \pm 0.3$ | $-1.04 \pm 1.59$ | |
| | 59736.28 | $2.13 \times 10^{29}$ | $2.29 \times 10^{29}$ | $4.68 \times 10^{28}$ | $-0.84 \pm 0.65$ | $-5.14 \pm 16.47$ | |
| 6478107251392112640 | 59776.63 | $6.40 \times 10^{29}$ | $1.40 \times 10^{30}$ | $1.81 \times 10^{30}$ | $0.71 \pm 0.19$ | $-0.33 \pm 0.58$ | 6 |
| | 58645.25 | $6.63 \times 10^{29}$ | $1.14 \times 10^{30}$ | $1.51 \times 10^{30}$ | $0.18 \pm 0.26$ | $-0.26 \pm 0.89$ | |
| | 58941.62 | $5.74 \times 10^{29}$ | $1.06 \times 10^{30}$ | $1.96 \times 10^{30}$ | $0.34 \pm 0.25$ | $0.61 \pm 0.72$ | |
| | 59382.16 | $3.80 \times 10^{29}$ | $5.12 \times 10^{29}$ | $7.59 \times 10^{29}$ | $-0.35 \pm 0.42$ | $0.03 \pm 1.49$ | |
| | 58808.69 | $3.49 \times 10^{29}$ | $4.57 \times 10^{29}$ | $3.02 \times 10^{28}$ | $-0.41 \pm 0.52$ | $-8.1 \pm 38.87$ | |
| | 59346.44 | $2.55 \times 10^{29}$ | $2.75 \times 10^{29}$ | $7.74 \times 10^{28}$ | $-0.84 \pm 0.69$ | $-4.31 \pm 13.19$ | |
| 6471134835842209792 | 58961.91 | $2.24 \times 10^{30}$ | $4.92 \times 10^{30}$ | $6.11 \times 10^{30}$ | $0.73 \pm 0.14$ | $-0.43 \pm 0.41$ | 5 |
| | 58655.80 | $7.54 \times 10^{29}$ | $1.20 \times 10^{30}$ | $2.20 \times 10^{30}$ | $0.02 \pm 0.47$ | $0.59 \pm 1.2$ | |
| | 58966.06 | $5.51 \times 10^{29}$ | $8.38 \times 10^{29}$ | $4.47 \times 10^{29}$ | $-0.08 \pm 0.71$ | $-2.64 \pm 6.81$ | |
| | 59887.43 | $7.24 \times 10^{29}$ | $7.66 \times 10^{29}$ | $1.45 \times 10^{30}$ | $-0.88 \pm 0.66$ | $0.66 \pm 2.08$ | |
| | 58591.69 | $6.20 \times 10^{29}$ | $6.88 \times 10^{29}$ | $8.49 \times 10^{28}$ | $-0.77 \pm 0.76$ | $-6.46 \pm 32.38$ | |
| 4743636248509810944 | 59526.67 | $8.00 \times 10^{29}$ | $1.36 \times 10^{30}$ | $1.17 \times 10^{30}$ | $0.15 \pm 0.14$ | $-1.38 \pm 0.7$ | 5 |
| | 59547.09 | $8.03 \times 10^{29}$ | $8.69 \times 10^{29}$ | $1.06 \times 10^{30}$ | $-0.83 \pm 0.19$ | $-0.47 \pm 0.83$ | |
| | 59379.72 | $2.37 \times 10^{29}$ | $2.78 \times 10^{29}$ | $3.38 \times 10^{29}$ | $-0.65 \pm 0.58$ | $-0.5 \pm 2.97$ | |
| | 59839.07 | $2.82 \times 10^{29}$ | $2.61 \times 10^{29}$ | $6.20 \times 10^{29}$ | $-1.17 \pm 0.57$ | $1.26 \pm 1.41$ | |
| | 59781.46 | $2.46 \times 10^{29}$ | $2.50 \times 10^{29}$ | $4.74 \times 10^{29}$ | $-0.97 \pm 0.63$ | $0.67 \pm 1.85$ | |
| 4742040410461492096 | 58705.72 | $2.27 \times 10^{28}$ | $3.87 \times 10^{28}$ | $8.66 \times 10^{28}$ | $0.17 \pm 0.22$ | $1.1 \pm 0.46$ | 5 |
| | 59373.10 | $1.64 \times 10^{28}$ | $3.34 \times 10^{28}$ | $7.71 \times 10^{28}$ | $0.56 \pm 0.26$ | $1.18 \pm 0.5$ | |
| | 58719.74 | $1.03 \times 10^{28}$ | $2.00 \times 10^{28}$ | $7.43 \times 10^{28}$ | $0.44 \pm 0.46$ | $2.43 \pm 0.6$ | |
| | 59741.04 | $1.37 \times 10^{28}$ | $1.84 \times 10^{28}$ | $7.67 \times 10^{27}$ | $-0.36 \pm 0.44$ | $-3.28 \pm 5.48$ | |
| | 59413.11 | $6.80 \times 10^{27}$ | $1.82 \times 10^{28}$ | $7.61 \times 10^{27}$ | $1.16 \pm 0.57$ | $-3.28 \pm 4.73$ | |
| 4946938113149426944 | 59158.65 | $4.11 \times 10^{27}$ | $9.07 \times 10^{27}$ | $1.74 \times 10^{28}$ | $0.74 \pm 0.08$ | $0.7 \pm 0.19$ | 4 |
| | 59109.20 | $2.44 \times 10^{27}$ | $7.05 \times 10^{27}$ | $1.53 \times 10^{28}$ | $1.32 \pm 0.13$ | $1.02 \pm 0.22$ | |
| | 59040.35 | $1.01 \times 10^{27}$ | $1.96 \times 10^{27}$ | $1.33 \times 10^{27}$ | $0.44 \pm 0.34$ | $-2.02 \pm 2.06$ | |
| | 59676.48 | $1.06 \times 10^{27}$ | $1.43 \times 10^{27}$ | $1.26 \times 10^{27}$ | $-0.36 \pm 0.39$ | $-1.34 \pm 2.27$ | |
| 6452221724017193216 | 59778.00 | $1.55 \times 10^{29}$ | $3.64 \times 10^{29}$ | $5.42 \times 10^{28}$ | $0.87 \pm 0.45$ | $-5.97 \pm 10.39$ | 3 |
| | 58667.89 | $1.39 \times 10^{29}$ | $2.92 \times 10^{29}$ | $4.16 \times 10^{29}$ | $0.61 \pm 0.51$ | $-0.07 \pm 1.37$ | |
| | 58958.14 | $1.62 \times 10^{29}$ | $2.69 \times 10^{29}$ | $4.46 \times 10^{29}$ | $0.11 \pm 0.5$ | $0.32 \pm 1.36$ | |
| 4910693055937045888 | 59834.56 | $7.41 \times 10^{28}$ | $1.40 \times 10^{29}$ | $2.70 \times 10^{29}$ | $0.39 \pm 0.44$ | $0.72 \pm 0.96$ | 3 |
| | 58949.56 | $5.83 \times 10^{28}$ | $1.35 \times 10^{29}$ | $2.93 \times 10^{29}$ | $0.83 \pm 0.5$ | $1.03 \pm 0.88$ | |
| | 59428.56 | $7.81 \times 10^{28}$ | $1.15 \times 10^{29}$ | $2.90 \times 10^{28}$ | $-0.15 \pm 0.46$ | $-4.6 \pm 8.6$ | |
| 4723685987981062656 | 58747.57 | $7.91 \times 10^{29}$ | $2.31 \times 10^{30}$ | $3.20 \times 10^{30}$ | $1.34 \pm 0.24$ | $-0.14 \pm 0.53$ | 3 |
| | 59739.27 | $4.52 \times 10^{29}$ | $1.02 \times 10^{30}$ | $2.96 \times 10^{30}$ | $0.79 \pm 0.47$ | $1.77 \pm 0.62$ | |
| | 59361.00 | $3.21 \times 10^{29}$ | $6.80 \times 10^{29}$ | $5.93 \times 10^{29}$ | $0.65 \pm 0.59$ | $-1.36 \pm 2.68$ | |
| 6404861806838482944 | 59872.95 | $1.25 \times 10^{30}$ | $2.50 \times 10^{30}$ | $2.83 \times 10^{30}$ | $0.52 \pm 0.32$ | $-0.67 \pm 1.06$ | 2 |
| | 59726.28 | $7.35 \times 10^{29}$ | $1.87 \times 10^{30}$ | $2.12 \times 10^{30}$ | $1.05 \pm 0.52$ | $-0.68 \pm 1.52$ | |
| 6396977135940047616 | 59338.88 | $1.80 \times 10^{28}$ | $2.64 \times 10^{28}$ | $3.30 \times 10^{28}$ | $-0.16 \pm 0.33$ | $-0.42 \pm 1.37$ | 2 |
| | 59393.69 | $8.83 \times 10^{27}$ | $1.85 \times 10^{28}$ | $4.51 \times 10^{28}$ | $0.62 \pm 0.55$ | $1.33 \pm 0.96$ | |
| 4979859106192681344 | 58594.72 | $3.44 \times 10^{28}$ | $7.56 \times 10^{28}$ | $1.27 \times 10^{29}$ | $0.72 \pm 0.56$ | $0.35 \pm 1.42$ | 2 |
| | 58808.36 | $2.93 \times 10^{28}$ | $7.16 \times 10^{28}$ | $5.35 \times 10^{27}$ | $0.96 \pm 0.63$ | $-7.78 \pm 30.45$ | |
| 4928347433108236416 | 59447.56 | $5.48 \times 10^{27}$ | $1.35 \times 10^{28}$ | $2.26 \times 10^{28}$ | $0.97 \pm 0.39$ | $0.36 \pm 0.85$ | 2 |
| | 58590.05 | $5.00 \times 10^{27}$ | $7.40 \times 10^{27}$ | $1.30 \times 10^{28}$ | $-0.14 \pm 0.63$ | $0.46 \pm 1.94$ | |
| 4726387007013852672 | 59507.78 | $5.84 \times 10^{27}$ | $1.67 \times 10^{28}$ | $4.38 \times 10^{28}$ | $1.3 \pm 0.09$ | $1.52 \pm 0.13$ | 2 |
| | 58982.09 | $1.35 \times 10^{27}$ | $2.37 \times 10^{27}$ | $1.47 \times 10^{27}$ | $0.23 \pm 0.49$ | $-2.26 \pm 3.39$ | |
| 4713771622913507328 | 58798.28 | $1.02 \times 10^{28}$ | $1.57 \times 10^{28}$ | $1.88 \times 10^{28}$ | $-0.06 \pm 0.21$ | $-0.53 \pm 0.79$ | 2 |
| | 59710.62 | $3.84 \times 10^{27}$ | $7.03 \times 10^{27}$ | $2.15 \times 10^{28}$ | $0.32 \pm 0.49$ | $1.92 \pm 0.72$ | |
| 4708842443566593536 | 59345.82 | $1.72 \times 10^{30}$ | $5.16 \times 10^{30}$ | $1.03 \times 10^{31}$ | $1.4 \pm 0.18$ | $0.8 \pm 0.32$ | 2 |
| | 59854.89 | $1.25 \times 10^{30}$ | $1.71 \times 10^{30}$ | $1.74 \times 10^{30}$ | $-0.32 \pm 0.39$ | $-0.96 \pm 1.7$ | |

## ORCID iDs

C. Tandoi https://orcid.org/0000-0002-2077-6004
S. Guns https://orcid.org/0000-0001-7143-2853
A. Foster https://orcid.org/0000-0002-7145-1824
P. A. R. Ade https://orcid.org/0000-0002-5127-0401
A. J. Anderson https://orcid.org/0000-0002-4435-4623
M. Archipley https://orcid.org/0000-0002-0517-9842
L. Balkenhol https://orcid.org/0000-0001-6899-1873
A. N. Bender https://orcid.org/0000-0001-5868-0748
B. A. Benson https://orcid.org/0000-0002-5108-6823
F. Bianchini https://orcid.org/0000-0003-4847-3483
L. E. Bleem https://orcid.org/0000-0001-7665-5079
F. R. Bouchet https://orcid.org/0000-0002-8051-2924
J. E. Carlstrom https://orcid.org/0000-0002-2044-7665
T. W. Cecil https://orcid.org/0000-0002-7019-5056
P. M. Chichura https://orcid.org/0000-0002-5397-9035
T. M. Crawford https://orcid.org/0000-0001-9000-5013
C. Daley https://orcid.org/0000-0002-3760-2086
D. Dutcher https://orcid.org/0000-0002-9962-2058
W. Everett https://orcid.org/0000-0002-5370-6651
K. R. Ferguson https://orcid.org/0000-0002-4928-8813
R. Gualtieri https://orcid.org/0000-0003-4245-2315
G. P. Holder https://orcid.org/0000-0002-0463-6394
J. C. Hood https://orcid.org/0000-0003-4157-4185
A. T. Lee https://orcid.org/0000-0003-3106-3218
M. Millea https://orcid.org/0000-0001-7317-0551
G. I. Noble https://orcid.org/0000-0002-5254-243X
Z. Pan https://orcid.org/0000-0002-6164-9861
K. A. Phadke https://orcid.org/0000-0001-7946-557X
A. Rahlin https://orcid.org/0000-0003-3953-1776
C. L. Reichardt https://orcid.org/0000-0003-2226-9169
C. Reuter https://orcid.org/0000-0001-7477-1586
J. A. Sobrin https://orcid.org/0000-0001-6155-5315
C. Umilta https://orcid.org/0000-0002-6805-6188
J. D. Vieira https://orcid.org/0000-0001-7192-3871
N. Whitehorn https://orcid.org/0000-0002-3157-0407
W. L. K. Wu https://orcid.org/0000-0001-5411-6920